\documentclass[11pt, a4paper, twoside, twocolumn, titlepage]{article}
\usepackage{graphicx}
\usepackage[normalem]{ulem} 
\usepackage{amsmath,amssymb}
\usepackage{xcolor}
\newcommand{\mathbfit}[1]{\textbf{\textit{#1}}}
\newcommand{\new}[1]{\textcolor{black}{#1}}

\title{\textbf{Highly ordered magnetic fields in the tail of the jellyfish galaxy JO206}}

\author{Ancla M\"uller\footnote{amueller@astro.rub.de, }$^,$\footnote{Ruhr University Bochum, Faculty of Physics and Astronomy, Astronomical Institute, Universit\"atsstr. 150, 44801 Bochum, Germany},
          Bianca Poggianti\footnote{INAF-Astronomical Observatory of Padova vicolo dell'Osservatorio 5, 35122 Padova, Italy}, Christoph Pfrommer\footnote{Leibniz-Institut f\"ur Astrophysik Potsdam (AIP), An der Sternwarte 16, D-14482 Potsdam, Germany}, Bj\"orn Adebahr$^2$,\\ Paolo Serra\footnote{INAF- Osservatorio Astronomico di Cagliari, Via della Scienza 5, I-09047 Selargius (CA), Italy}, Alessandro Ignesti\footnote{Dipartimento di Fisica e Astronomia, Università di Bologna, Via Gobetti 93/2, 40129 Bologna, Italy}$^,$\footnote{INAF, Istituto di Radioastronomia di Bologna, Via Gobetti 101, 40129 Bologna, Italy}, Martin Sparre$^{4,}$\footnote{Institut f\"ur Physik und Astronomie, Universit\"at Potsdam, Karl-Liebknecht-Str.24/25, 14476 Golm, Germany}, Myriam Gitti$^{6,7}$,\\ Ralf-J\"urgen Dettmar$^2$, Benedetta Vulcani$^3$, Alessia Moretti$^3$}

\begin{document}
\maketitle
%

\textbf{Jellyfish galaxies have long tails of gas that is stripped from the disc by ram pressure \new{due to the motion of galaxies in the intracluster medium} in galaxy clusters. We present the first measurement of the magnetic field strength and orientation within the disc and the (90\,kpc-long) \new{$\rm H\alpha$-emitting} tail of the jellyfish galaxy JO206. The tail has a large-scale magnetic field ($>4.1\,\mu$G), a steep radio spectral index ($\alpha \sim -2.0$), indicating an aging of the electrons propagating away from the star-forming regions, and extremely high fractional polarisation ($>50$\,\%), indicating low turbulent motions. \new{The magnetic field vectors are aligned with (parallel to) the direction of the ionised-gas tail and stripping direction.}
\new{High-resolution simulations of a large, cold gas cloud that is exposed to} a hot, magnetised turbulent wind show that the high fractional polarisation and the ordered magnetic field can be explained by accretion of draped magnetised plasma from the hot wind that condenses onto the external layers of the tail, where it is adiabatically compressed and sheared. The ordered magnetic field, preventing heat and momentum exchange, may be a key factor in allowing in-situ star formation in the tail.
}
%
%

The star-formation rate and gas content of galaxies are strongly correlated \cite{Schmidt,Kennicutt1989}.
In order to understand the full picture of galaxy evolution, it is necessary to grasp the role of gas inflows and outflows from galaxies.
Among several processes that can remove gas from galaxies, 
the pressure exerted by the hot intracluster medium (ICM) on the interstellar medium of orbiting galaxies can efficiently remove gas via "ram-pressure stripping" in galaxy clusters and, sometimes, groups \cite{Cayatte,Kenney,Jaffe,Williams,Montenegro,Serra}. 

Recently, extreme ram-pressure stripped galaxies with long gaseous tails, named jellyfish galaxies, are found moving at high speed in clusters \cite{Jaffe,Smith2010,Fumagalli2014}.
In most cases, significant on-going star formation is detected within the galactic disc and far out in the tails \cite{Poggianti19, Vulcani}, demonstrating that gas stripping is not associated with an immediate quenching of star formation.
Among the most compelling open questions are the origin of the in-situ star formation in ram-pressure stripped tails, and whether magnetic fields enable star-formation activity in jellyfish tails.

Magnetic fields can play an active role in star-formation scenarios.
On parsec scales molecular clouds collapse by magnetic fields transporting angular momentum \cite{Crutcher}. 
Spiral galaxies show large-scale magnetic fields along and between their optical spiral arms and in their halos \cite{Beck13} anchoring the magnetic field in the molecular cloud complexes \cite{Li}. 
On kilo-parsec scales the flow of diffuse interstellar gas can be controlled by magnetic forces of barred galaxies \cite{Beck05II}.
So far, a few magnetic fields of moderately ram-pressure stripped galaxies have been observed in the close vicinity of galaxy discs \new{\cite{Gavazzi88,Gavazzi95,Vollmer10,Vollmer13}}, but no study was conducted on the long tails of jellyfish galaxies. 

Several simulations have been performed of ram-pressure stripped galaxies including magnetic fields. They find that 1) the reduced compressibility of magnetised plasma makes \new{narrower tails, which are smoother,} less clumpy and filamentary-like in comparison to the non-magnetised case \new{\cite{Vijayaraghavan2017,Martinez}}, 2) the stripping rate of the disc gas is not significantly affected by the magnetic field strength and morphology of the galaxy \cite{Tonnesen}, and 3) asymmetric polarised ridges can be explained through adiabatic compression of galaxy magnetic fields at the leading edge \cite{Vollmer07}. Alternatively, they can be explained by galaxies sweeping up the intracluster magnetic field as they orbit inside a galaxy cluster developing a magnetic draping sheath \cite{Dursi}, which is lit up with cosmic rays from the galaxies’ supernovae, generating coherent polarised emission at the galaxies’ leading edges \cite{Pfrommer10}. Simulated galaxies that interact with a magnetised ICM show highly ordered magnetic fields in the tails \cite{Ruszkowski} suppressing strong Kelvin-Helmholtz instabilities at the interface between the stripped gas and the ICM \cite{Berlok}.

\section*{Observational results}
We combine a new 50.5\,h JVLA C-array S-band (2.7\,GHz) observation with a previous 20\,h JVLA C-array L-band (1.4\,GHz) observation \cite{Ramatsoku} (see Methods)
which results in the first detection of magnetic field orientation in the 90\,kpc-long \new{H$\alpha$-emitting} tail of stripped gas of the galaxy JO206 selected from the GAs Stripping Phenomena in galaxies (GASP) survey \cite{Poggianti17}. JO206 is a massive galaxy (stellar mass $M_\star \sim 9 \times 10^{10}$\,M$_\odot$), hosts an active galactic nucleus (AGN), and undergoes extreme ram-pressure stripping in the low mass cluster IIZW108 \cite{Poggianti17,Poggianti17n} (see also SI, Sect. 1).

\begin{figure*}[t]
    \begin{minipage}[t]{.5\textwidth}
        \centering
        \includegraphics[width=\textwidth]{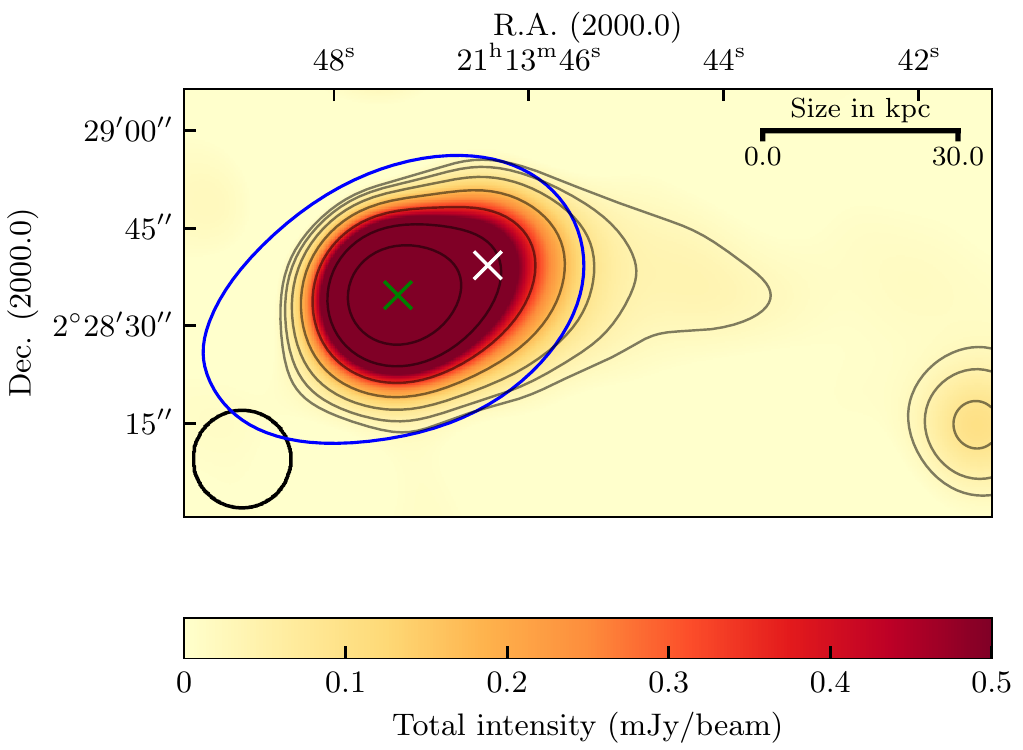}
    \end{minipage}  
    \hfill
    \begin{minipage}[t]{.5\textwidth}
        \centering
        \includegraphics[width=\textwidth]{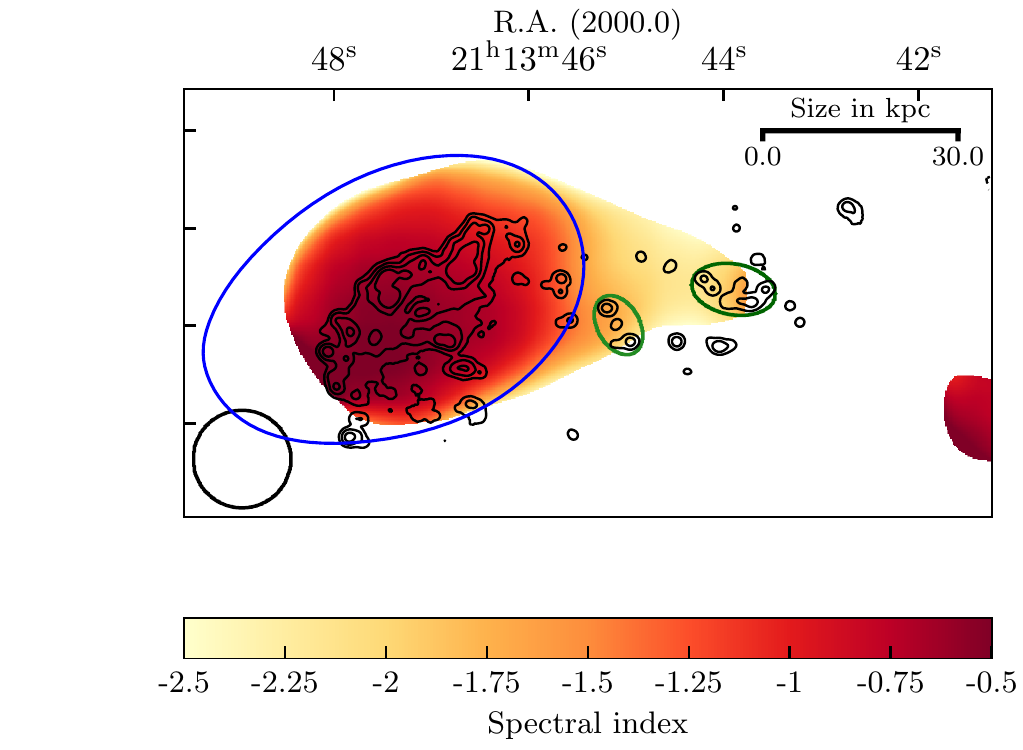}
    \end{minipage} 
     \begin{minipage}[t]{.5\textwidth}
        \centering
        \includegraphics[width=\textwidth]{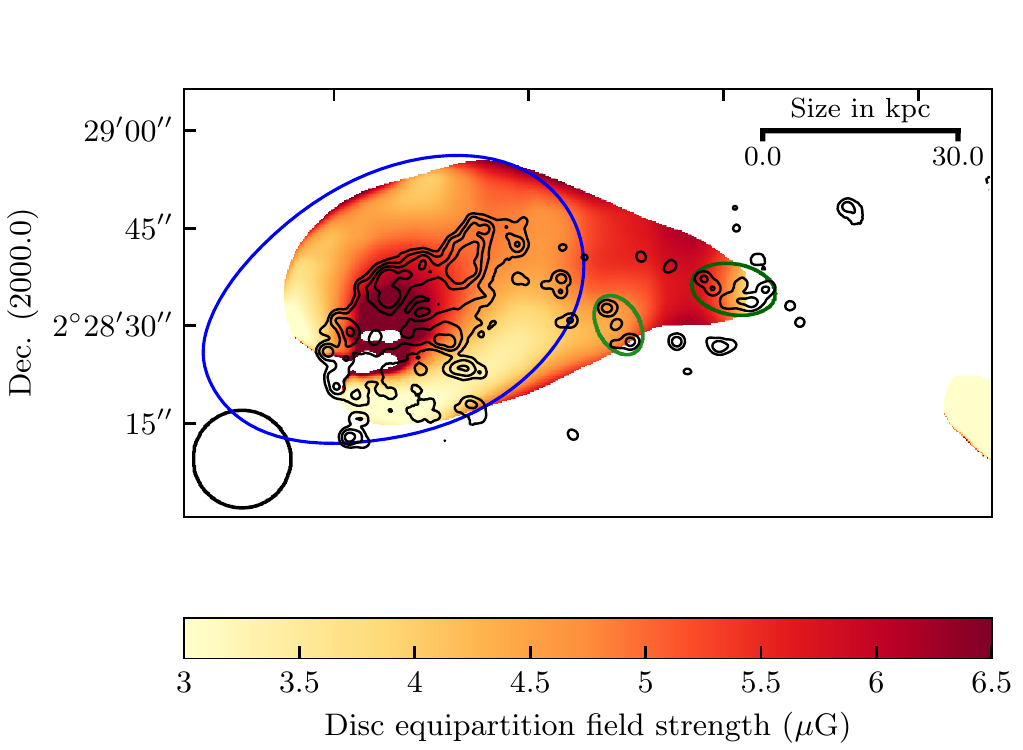}
    \end{minipage}  
    \hfill
    \begin{minipage}[t]{.5\textwidth}
        \centering
        \includegraphics[width=\textwidth]{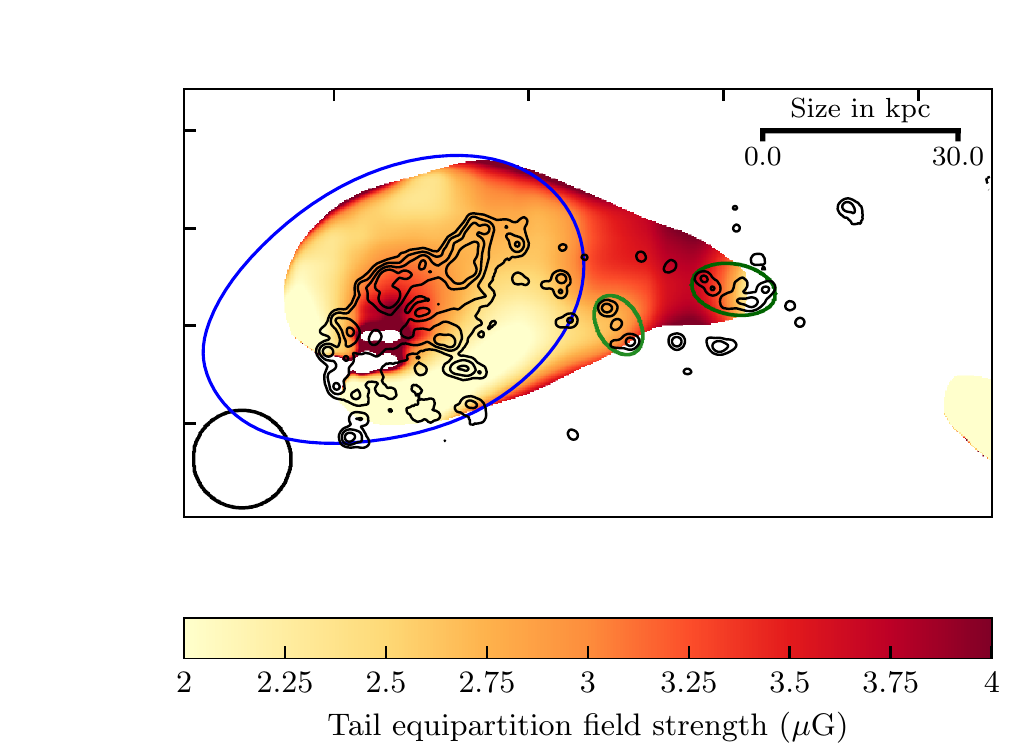}
    \end{minipage}
    \caption{Total intensity results of the 2.7\,GHz data. The blue contour defines the optical stellar disk convolved with the 2.7\,GHz resolution. The white area contains no information. \textbf{Top left}: Thermal-corrected 2.7\,GHz total intensity map with superimposed contour levels of $\epsilon \times (1,\;2,\;4,\;8,\;16,\;32,\;64,\;128,\;256$) with $\epsilon = 21.5\,\mu$Jy/beam corresponding to 3\,$\sigma$ and a green and white cross at the peak emission of the total and polarized intensity, respectively.
    The total intensity is corrected for the thermal contribution as described in Section 1 (see SI). \textbf{Top right}: Derived thermal corrected spectral index map based on the 2.7\,GHz and 1.4\,GHz map. \textbf{Bottom left}: Derived lower limits of the disc equipartition field strength computed with a pathlength of 2.8\,kpc. In the small white regions the field strength cannot be estimated due to the flat spectral index ($> -0.54$). Such flat spectra can be caused by synchrotron self-absorption by an AGN, or thermal absorption in star-forming regions \cite{Condon,Kaiser}. \textbf{Bottom right}: Derived lower limits of the tail equipartition field strength computed with a pathlength of 28.5\,kpc.
    The contour levels of $\epsilon \times (1,\;3,\;9,\;27,\;81,\;283$) with $\epsilon = 1\cdot 10^{-19}\,$erg/s/cm$^2$ in H$\alpha$ are superimposed to the spectral index and magnetic field maps, and two star forming complexes are marked by a green and dark green ellipsoid.} \label{fig:Results}
\end{figure*}

The 2.7\,GHz total intensity map is presented in Figure \ref{fig:Results} (top left).
The integrated total flux density of JO206 is ($3.60\pm0.18$)\,mJy.
It reaches ($3.5\pm0.2$)\,mJy in the disc (defined by the blue contour in Fig. \ref{fig:Results} corresponding to the stellar disc \cite{Marco} convolved to the 2.7\,GHz resolution) and ($0.10\pm 0.01$)\,mJy in the tail (defined by the outermost radio contour, which is $3\,\sigma$ for total intensity, $4.5\,\sigma$ for polarisation, excluding the disc region).

The peak emission (green cross in Fig. \ref{fig:Results}, top left) is \new{co-located with the AGN (central source, \new{see SI, Sect. 8}) contributing to the disc total flux density.}
We find a projected tail-length of $\sim 40$\,kpc in total intensity, which morphology is in good agreement with the H$\alpha$ emission, though the H$\alpha$ extends even further (Fig. \ref{fig:Results}, black contours).

The observed spectral index map at 2.7\,GHz and 1.4\,GHz \new{(assuming $I_{\nu}\propto \nu^\alpha$ where $I_{\nu}$ is the synchrotron surface brightness at frequency $\nu$)} is presented in Figure \ref{fig:Results} (top right). The integrated spectral index $\alpha$ reaches $-0.74\pm0.09$ (disc) and $-2.04\pm0.09$ (tail).
The disc spectral index is slightly steeper than the injection spectral index for star-forming regions ($\alpha\approx -0.6$ \cite{Niklas}), however consistent within the uncertainties, given our limited resolution.

The steep spectral index in the jellyfish tail most likely indicates an aging of the electrons \cite{Liu,Winner} that were produced in supernovae in star-forming regions and lost their energy (aged) during their propagation away from their origin being consistent with recent findings \cite{Chen}. 

We assume equipartition between the magnetic field and cosmic-ray energy density to estimate the magnetic field strength (using \new{Equation 3 in} \cite{Beck05} 
and a cosmic-ray proton-to-electron number density ratio $K_0$ of 100).
\new{Additionally, we assume different pathlengths for the disc (2.2\,kpc and 2.8\,kpc) and tail (20\,kpc and 28.5\,kpc) corresponding to a thick disc (frequency dependent scale height \cite{Krause}) and the tail being cylindrical (see SI, Sect. 3).}
We find an integrated disc magnetic field of \new{6.7\,$\mu$G to 7.1\,$\mu$G} and estimate a lower limit of 4.1\,$\mu$G in the tail. 
The lower limit results of the disc and tail are shown in Figure \ref{fig:Results}, bottom left and right, respectively.
The disc magnetic field is comparable to that observed in star-forming spiral galaxies \cite{Beck09}.

Because the synchrotron cooling time depends on the magnetic field strength, it can also be estimated from the tail length and galaxy velocity (see SI, Sect. 4). Reasonable assumptions lead to a tail magnetic field of \new{$B\approx2~\mu$G, which is consistent with the equipartition-derived value within the uncertainties.
Because JO206 is moving super-Alfv\'{e}nically with an Alfv\'{e}n Mach number of $\sim9$ we can estimate the magnetic field strength at the stagnation point (where the galaxy hits the ICM) to $\sim18~\mu$G (see SI, Sect. 5). Simulations show that the magnetic field strength decreases from the stagnation point along the galaxy ($B \approx (5-8)\,\mu$G) and predict a tail magnetic field of $B\approx(1-4)\,\mu$G that survives at least 100\,kpc away from the disc (see also SI, Sect. 9)
\cite{Dursi,Ruszkowski,Sparre2020}.}

\new{While each of the three methods for estimating magnetic field strength suffers from different systematic uncertainties, they all yield a consistent magnetic field in the tail of $B\approx(2-4)\,\mu$G. While the equipartition method suffers from the assumption of the path length, its restrictions on the spectral index (undefined for $\alpha > -0.54$, uncertain for $\alpha < -1.0$), and $K_0$ (only determined at the Solar radius), the electron cooling length argument assumes a constant gas velocity and magnetic field, depends on the uncertain inclination of the galaxy velocity, and neglects Fermi-II re-acceleration. The magnetic field estimate that relates at the field in the tail to the stagnation point suffers from theoretical uncertainties of this dynamical map because the magnetic field in the tail is set by a combination of velocity shear, compression due to gas condensation, and potentially a small-scale dynamo (for more details see SI, Sect. 3--5).}

\begin{figure*}[t]
    \begin{minipage}[t]{.5\textwidth}
        \centering
        \includegraphics[width=\textwidth]{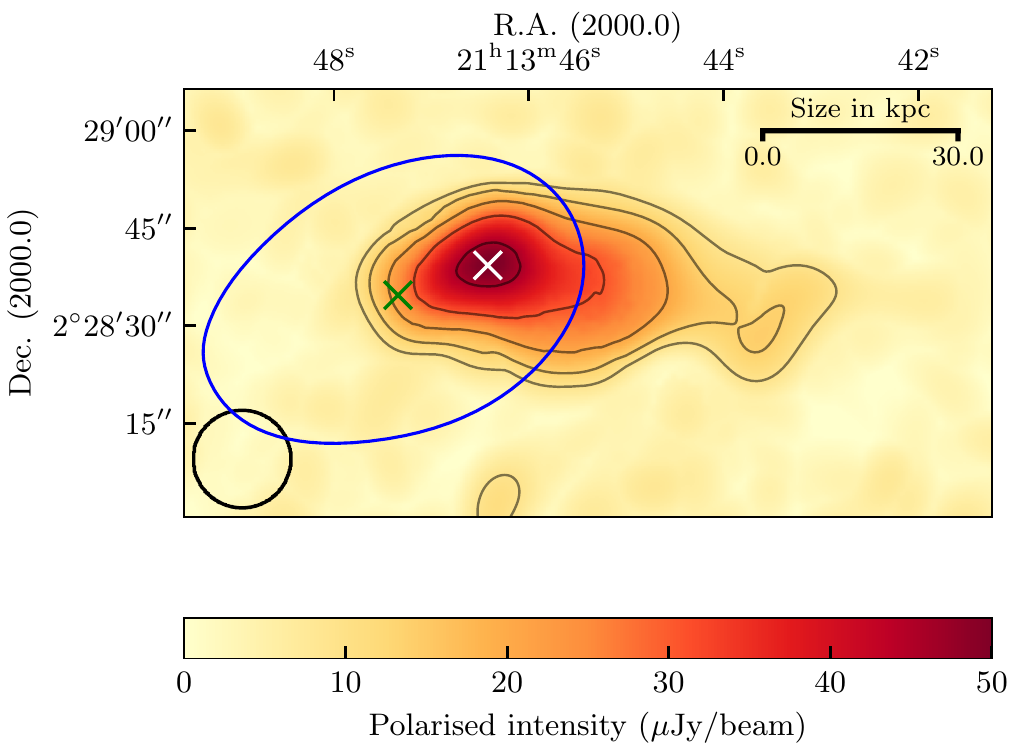}
    \end{minipage}  
    \hfill
    \begin{minipage}[t]{.5\textwidth}
        \centering
        \includegraphics[width=\textwidth]{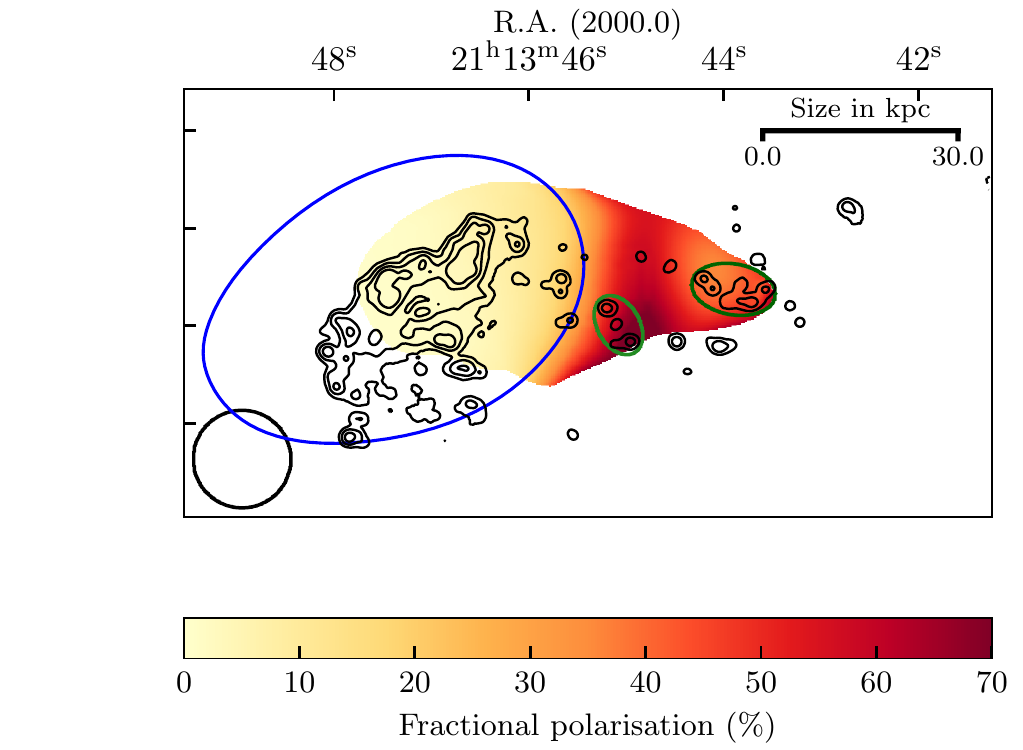}
    \end{minipage} 
    \caption{Polarization results of the 2.7\,GHz data. The blue contour defines the optical stellar disk convolved with the 2.7\,GHz resolution. The white area contains no information. \textbf{Left}: Polarized intensity map with superimposed contour levels of $\epsilon \times (1,\;1.5,\;2.25,\;3.375,\;5.0625$) with $\epsilon = 9\,\mu$Jy/beam corresponding to 4.5\,$\sigma$ and a green and white cross at the peak emission of the total and polarized intensity, respectively. \textbf{Right}: Derived fractional polarization map. The contour levels of $\epsilon \times (1,\;3,\;9,\;27,\;81,\;283$) with $\epsilon = 1\cdot 10^{-19}\,$erg/s/cm$^2$ in H$\alpha$ are superimposed and two star forming complexes are marked by a green and dark green ellipsoid.
    } \label{fig:Results2}
\end{figure*}

We now examine the polarised intensity of JO206 (Fig.\ \ref{fig:Results2}, left panel). 
The integrated polarised intensity is ($0.136\pm 0.007$)\,mJy, while it reaches ($0.083\pm 0.004$)\,mJy (disc) and ($0.054\pm 0.003$)\,mJy (tail), accounting for a bias correction (see Methods and SI, Sect. 6). 
For the first time, we find polarised intensity far in the jellyfish tail. The extension of polarised emission in the tail is comparable to the total intensity (Fig. \ref{fig:Results}, top left) while we see no polarised emission in the northern, eastern, and southern parts of the disc. \new{We also do not detect polarised emission in a higher resolved map, likely due to beam depolarisation effects (see SI, Sect. 8)}.

The peak intensity of the polarised emission is located at the western side of the disc (white cross in Fig.\ \ref{fig:Results2}, left, \new{see also SI, Sect. 8}), and does not coincide with the total intensity distribution (green cross).
The polarised emission peak is co-located with the peak intensity in HI and secondary peaks of the H$\alpha$ and CO(2-1) emission maps within the uncertainties of the different resolutions (\cite{Ramatsoku}, Moretti et al. in prep.). While the total intensity peak seems to be \new{co-located with} the central source, the polarised intensity indicates a strong spatial correlation with the most active star-forming regions in the disc. 

The fractional polarisation, polarised intensity divided by the total intensity, is shown in Figure \ref{fig:Results2} (right).
We find the integrated fractional polarisation to be much higher in the tail ($54\pm 10$)\,\% than in the disc ($2\pm 6$)\,\%.
The fractional polarisation increases rapidly in the tail near the disc, from $\sim 10$\,\% to $\sim 30$\,\% (Fig.~\ref{fig:Results2}, right). While the polarised emission is uniformly decreasing far into the tail, the total intensity drops very fast in that particular region.

The disc values are comparable to those found in ram-pressure stripped Virgo cluster galaxies \cite{Vollmer10,Vollmer13}. Only sub-regions in star-forming spiral galaxies, \new{in particular the inter-arm regions}, show a fractional polarisation up to 50\,\% \cite{Beck15}, while their integrated fractional polarisation (disc) reaches between 0.5\,\% and 17.4\,\% \cite{Stil}.
Our high tail values are comparable to those in so-called head-tail galaxies, hosting an AGN, producing extended jets, moving inwards into a galaxy cluster, experiencing ram pressure, and showing fractional polarisation values of 60\,\% \cite{Miley}.

Thus, this is the first observation of such high fractional polarisation values in a star-forming galaxy undergoing ram-pressure stripping. 
Because the fractional polarisation is a measure for the uniformity of the magnetic field (a higher fractional polarisation corresponds to a lower turbulent motion),
this suggests an increased alignment of the tail magnetic field, consistent with the direct observation of the strong alignment.

\begin{figure*}[t]
    \centering
    \includegraphics[scale=1.4]{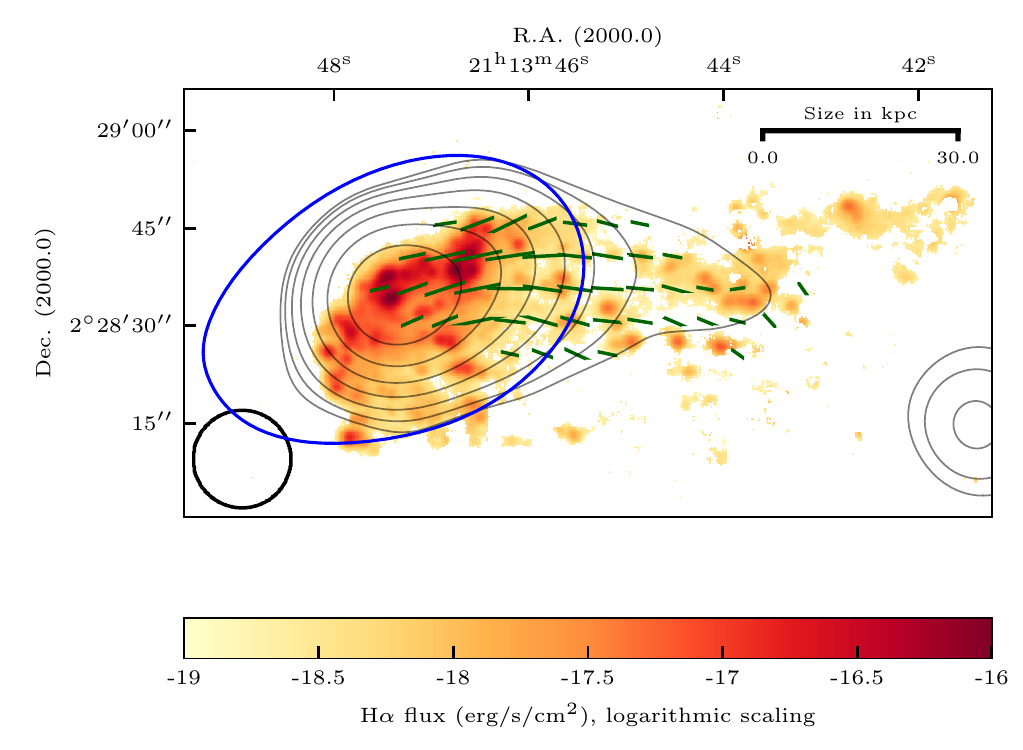}
    \caption{Large-scale alignment of the magnetic field vectors along the galaxy tail derived from the polarization angle. The magnetic field vectors are weighted by $10^{-5} \times$\,(polarized intensity) and shown in green. They are superimposed over the H$\alpha$ emission, which is shown in colour. The 2.7\,GHz total intensity contours are shown in black for comparison. The blue contour defines the optical stellar disc convolved with the 2.7\,GHz resolution. The white area contains no information.
    } \label{fig:Results3}
\end{figure*}

From the polarisation angle we calculate the intrinsic magnetic field configuration (Fig. \ref{fig:Results3}).
The length of each magnetic field vector corresponds to the polarised intensity weighted by $10^5\,\mu$Jy.
We find this magnetic field component aligned from the south-western part of the disc along the length of the jellyfish tail in stripping direction. 
From the magnetic field configuration, together with the fractional polarisation results, we find a highly ordered magnetic field in the jellyfish tail, in spite of the presence of star-forming regions (see H$\alpha$ emission in Fig. \ref{fig:Results3}). 
The presence of such \new{a large-scale} ordered magnetic field might support the star formation in the tail (see Discussion).

Because star formation triggers turbulent motion, it is appropriate to ask whether the star formation in the tail has a significant influence on \new{the spectral index (freshly induced electrons) and the fractional polarisation (higher turbulence).} We analyse two specific regions marked in green, one closer to the disc (inner) and the other farther out in the tail (outer, see. Figs. \ref{fig:Results} and \ref{fig:Results2}.
These regions enclose the strongest star-forming clumps in the tail, as visible from the H$\alpha$ contours showing a significantly higher star-formation-rate surface density (SFRD) than in their surroundings:
$\sim 10^{-2}$\,M$_\odot$/yr/kpc$^2$ (\new{inner}) and $\sim 10^{-2.2}$\,M$_\odot$/yr/kpc$^2$ (\new{outer}) versus
$\textrm{SFRD}\approx 10^{-4}$\,M$_\odot$/yr/kpc$^2$ in the surroundings (see Fig. 9 in \cite{Ramatsoku}).

These regions of SFRD maxima correspond to \new{maxima} of the radio spectral index $\alpha$ (Fig. \ref{fig:Results}, top right) of $-1.80\pm 0.02$ (\new{inner}) and $-1.94\pm 0.02$ (\new{outer}) in comparison to the mean of $-2.24\pm0.01$ in between the green areas of the tail. \new{This can clearly be seen in the outer region (electrons farther outside should be aged even more), while the inner one is embedded in an elongated morphology} both indicating a significant local contribution of non-aged electrons originating from in-situ star-forming regions also resulting in reduced equipartition field (Fig. \ref{fig:Results}, bottom right). 
The outer region also shows a reduced mean fractional polarisation (($43\pm6$)\,\% in comparison to ($56\pm6$)\,\% in between the green areas) indicating a higher turbulent component than in other regions of the tail (Fig. \ref{fig:Results2}, right), as expected if some turbulent motion is induced by local star formation.
\new{We cannot examine a similar behavior for the inner region.}   
Still, the contribution of the local in-situ star formation in the tail is not affecting the large-scale ordered magnetic field along the tail (Fig. \ref{fig:Results3}). \new{Because the H$\alpha$ star-forming regions are small in comparison to our resolution we do not expect that the supernovae-generated turbulence in the star-forming regions tangles the field on scales of our resolution.}

We conclude that the tail is composed of both turbulent but small star-forming regions where the spectral index is flatter (recently accelerated electrons), and regions of steeper spectral indices (aged electrons) and a highly large-scale ordered magnetic field \new{that} may prevent heat and momentum exchange, providing a suitable environment for star formation to take place.

\section*{Simulations}
\label{sec:Simulations}

To understand our observational findings we model the following scenario: after cosmic-ray electrons are accelerated \new{at supernova remnant shocks} in the galactic disc and the star-forming tail, they propagate into the ambient interstellar medium. As this medium experiences ram-pressure stripping, the cosmic rays are carried alongside and cool via synchrotron emission and inverse Compton interactions with cosmic microwave background photons. We model and compare the radio emission to the observation presented above. 

\new{We intentionally adopt a transparent simulation model of a galaxy-sized, cold-dense spherical cloud that is interacting with a hot-diffuse, supersonic wind. This enables us to understand the essential physics by varying the model (hydrodynamics versus magneto-hydrodynamics with different field topologies in wind and cloud) and parameters (different cloud sizes, wind velocities, and cloud-wind density contrasts). While those results are published elsewhere \cite{Sparre2020}, here we model the synchrotron emission of a subset of simulations and provide arguments in support of this model to capture the essential physics of a ram-pressure stripped galaxy (see SI).}

We perform three-dimensional magneto-hydrodynamic simulations \new{of this setup} with the moving mesh code AREPO \cite{Springel,Pakmor13,Pakmor15}. The cloud is a factor of $10^3$ denser than the wind and is initially in pressure equilibrium with the surroundings. We follow the radiative cooling of the gas with a standard cooling function of Solar metallicity and adopt a weak magnetic field: we assume a thermal-to-magnetic pressure ratio of ten (see \new{Methods}). 

\begin{figure*}[th!]
        \centering
        \includegraphics[width=0.85\textwidth]{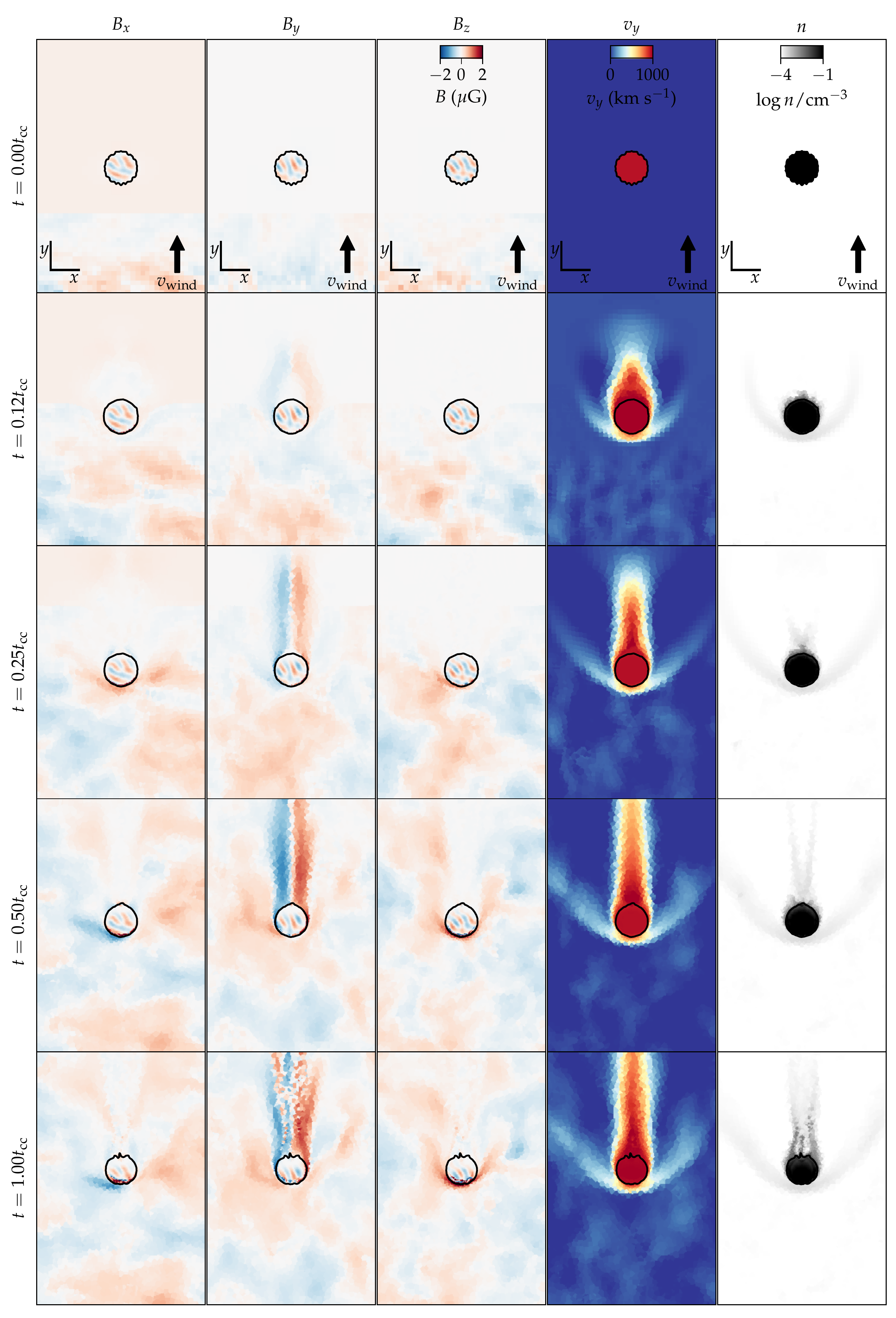}
        \caption{Initial time evolution of magnetic field components, the vertical velocity and the gas density in the simulation with a turbulent wind. This demonstrates the formation of a magnetic draping layer that enables condensation and accretion of hot ICM onto the tail and hence magnetic field alignment with the filamentary tail. Each panel measures 300 kpc $\times$ 450 kpc.}
        \label{fig:time_sequence}
\end{figure*}

We compare two simulations, one with magnetised turbulent wind and one with a homogeneous magnetic field that is oriented perpendicular to the cloud velocity (the field component that experiences magnetic draping, see \new{Figure~9 in the SI of \cite{Pfrommer10}, which shows draping of an inclined homogeneous magnetic field}). The turbulent and the ordered ICM magnetic field (upstream of the cloud) experience a change of their magnetic topology \new{downstream the cloud} as a result of the cloud-wind interaction. \new{Figure~\ref{fig:time_sequence} shows the initial cloud evolution in a turbulent wind.
An object moving through the magnetised ICM sweeps up turbulent magnetic field to build up a dynamically important sheath around the head (visible in $B_x$ and $B_z$) -- an effect called magnetic draping \cite{Dursi, Pfrommer10}. Simultaneously, a magnetised dense tail forms via condensation of the ICM \cite{Sparre2020}, thereby stretching and aligning the magnetic field with the tail and increasing $B_y$. Velocity shear of the ambient hot ICM induces vorticity and causes the cold-dense tail to continuously slow down.}

Which of the two \new{magnetic wind} configurations (if any at all) is able to reproduce the observations of an ordered tail magnetic field? To this end, we emulate the synchrotron observables at 2.7\,GHz, convolve the images with a Gaussian beam of size of the cloud radius (which is close to the synthesised beam of our observations), and impose a flux limit. 

To model cosmic-ray electron transport, we identify gas cells that have been part of the cold galactic gas (where cosmic rays are accelerated) and account for the cooling as they are advected with the wind. To this end, we assume a cosmic-ray electron energy distribution $f(E)\propto E^{-\alpha_e} \exp[-r/r_{\mathrm{cool}}]$, where  $\alpha_e=3$, $r$ is the distance from the \new{rear} of the cloud \new{and $r_{\mathrm{cool}}$ is the cooling length that is set by synchrotron and inverse Compton cooling (see SI).} We follow Lagrangian trajectories in our simulation and include the contribution of a gas cell to the synchrotron signal only if its trajectory emerged from the gas cloud at the initial time. For large clouds, the cold phase in the ram-pressure stripped tail is not destroyed by hydrodynamic instabilities but instead can accrete gas from the hot wind by mixing it into a layer with intermediate temperatures ($T\sim3\times10^5$\,K), from where it quickly cools and grows the cold tail \cite{Gronke,Li2020,Sparre2020}. 

\begin{figure*}[t]
        \includegraphics[width=\textwidth]{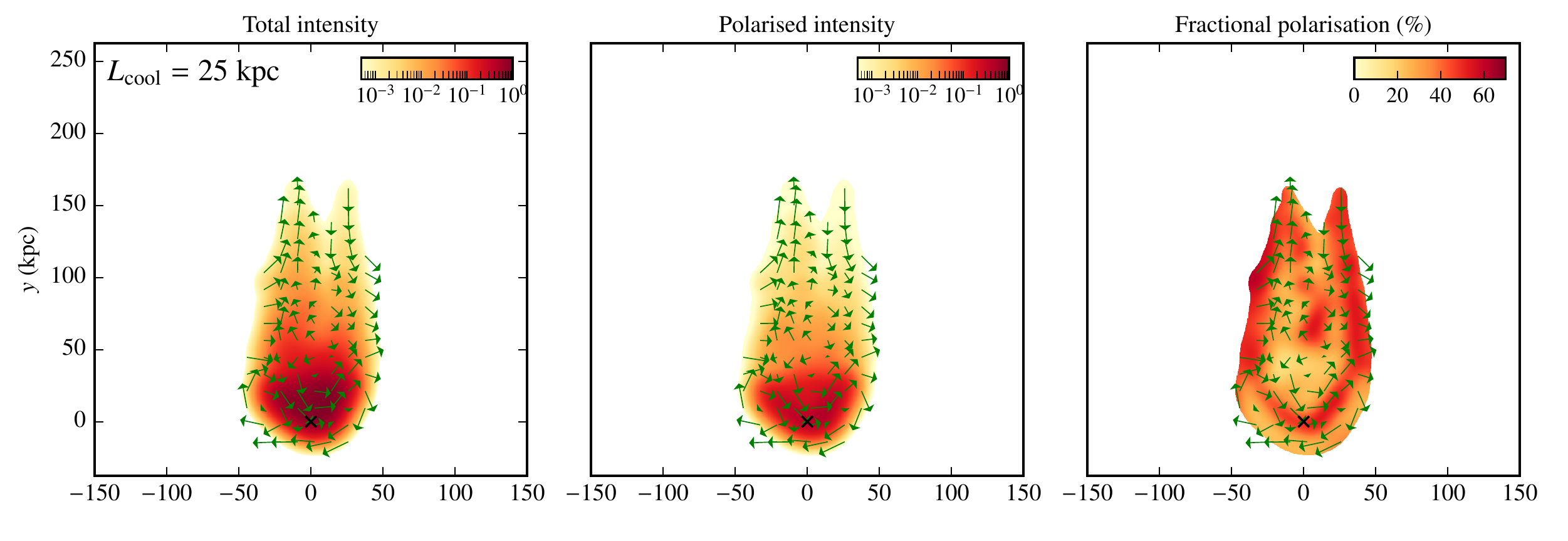}
        \includegraphics[width=\textwidth]{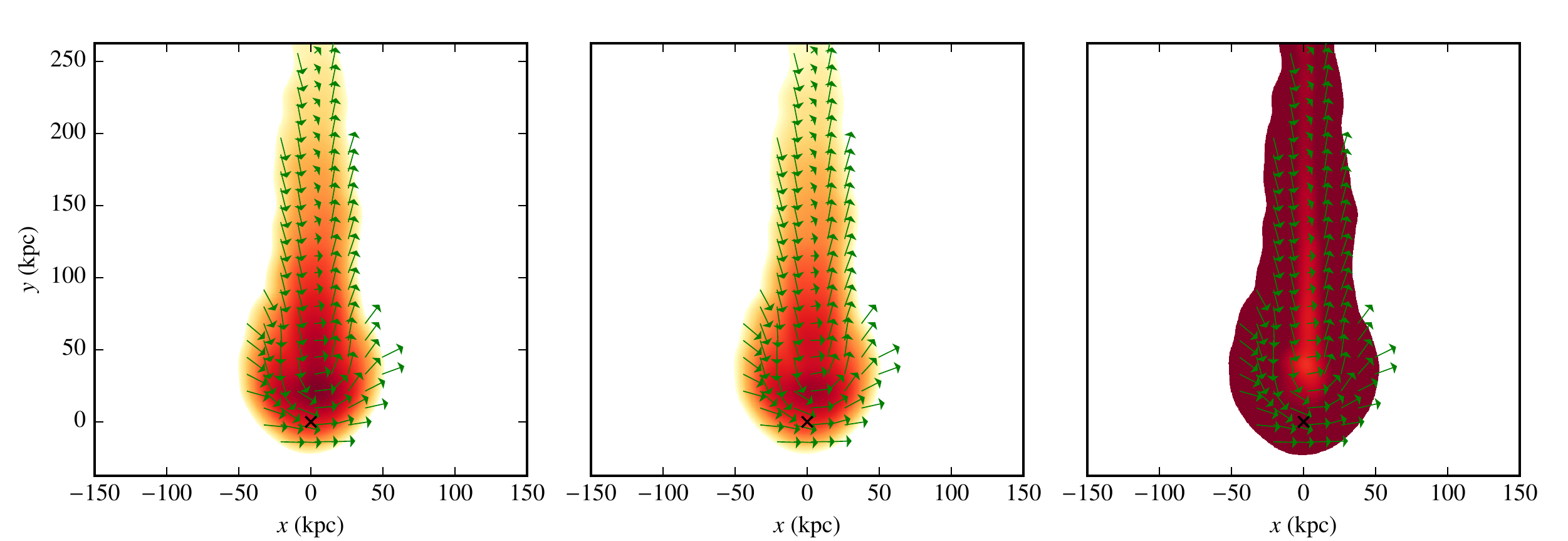}
    \caption{Simulated mock synchrotron observables (total intensity, polarized intensity, both normalised to the maximum total intensity, and fractional polarisation at $2.7$~GHz) of a cold ($T=10^4$~K) gas cloud that interacts with a supersonic ICM at four cloud crushing time scales. The green arrows show the magnetic field orientations and all images have been convolved with a Gaussian beam of $\mathrm{FWHM}=r_{\mathrm{cloud}} =19$~kpc. \textbf{The top panels} show a cloud that encounters a turbulent wind magnetic field, which gets aligned with the tail as a result of th cloud-wind interaction. In the \textbf{the bottom panels}, we show a cloud that encounters a homogeneous wind magnetic field that withstands alignment and produces a fractional polarisation that is larger than our observations, ruling out the combination of field topology and viewing angle for our observed jellyfish galaxy.} \label{fig:Simulations}
\end{figure*}

Figure~\ref{fig:Simulations} shows the cloud interacting with a turbulent wind (top panels) and a homogeneous magnetic field in the wind (bottom panels) that is initially oriented in the $x$ direction. We show the simulation at time $t=\new{4}\,t_{\mathrm{cc}}$, where
\begin{align}
t_\text{cc} \equiv \frac{r_\text{cloud}}{v_\text{wind}} \sqrt{\frac{n_\text{cloud}}{n_\text{wind}}}
\label{tcc}
\end{align}
is the cloud crushing time-scale in the absence of radiative cooling, and $n_\text{cloud}$ and $n_\text{wind}$ are the cloud and wind density, respectively. \new{For our cloud radius $r_\text{cloud}=25~\mathrm{kpc}$, wind velocity $v_\text{wind}=958~\mathrm{km~s}^{-1}$, and $n_\text{cloud}/n_\text{wind}=10^3$, we obtain $t_\text{cc} =8\times10^8~\mathrm{yr}$.} At this time, the tail of the cloud is starting to grow mass from the ICM \new{and the synchrotron emission extends over $\sim100$~kpc after 3.2~Gyr. In both simulations, we observe an ordered tail magnetic field that is aligned with the tail.} The reason for this ordered magnetic field is twofold. 1. Magnetic draping of the magnetic ICM field (see the bottom panels of Figure~\ref{fig:time_sequence}, also refs.~\cite{Dursi,Pfrommer10}). 2. Accretion of gas from the magnetic draping layer in the hot wind that condenses onto the tail causes the magnetic field to be adiabatically compressed and sheared by the differential velocity of the cold tail and wind. This explains the ordered magnetic field and hence, the high degree of fractional polarisation (see right-hand panels of Figure~\ref{fig:Simulations}). 

\begin{figure*}[t]
        \includegraphics[width=0.3725\textwidth]{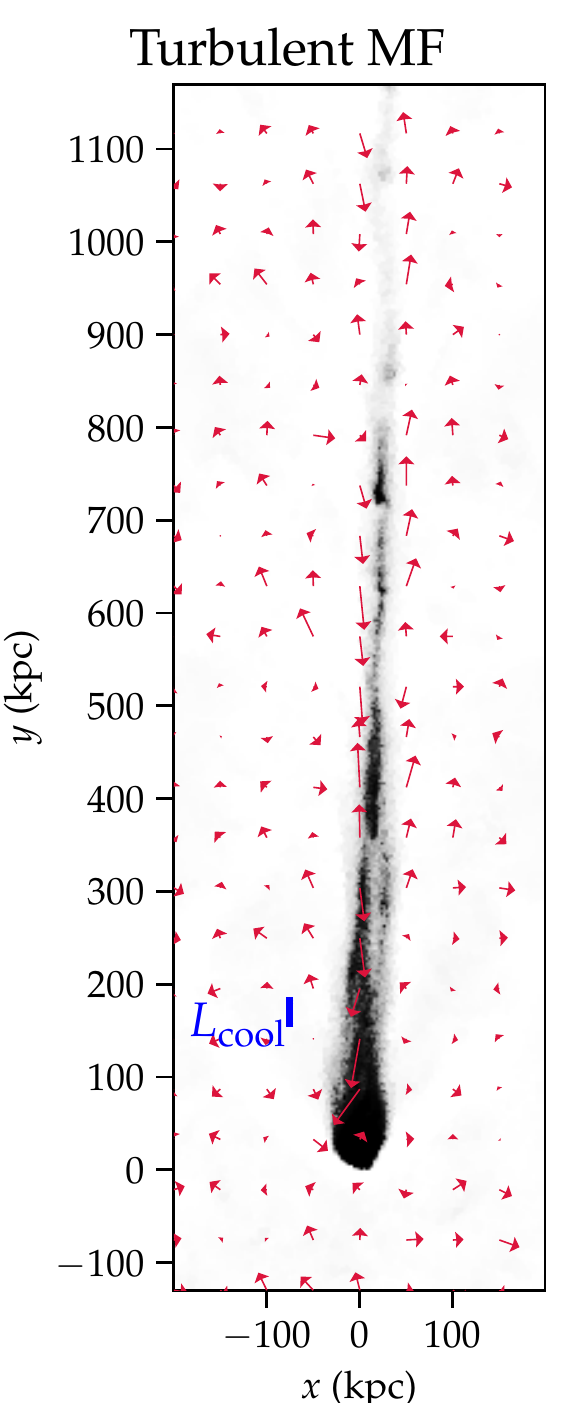}
        \includegraphics[width=0.275\textwidth]{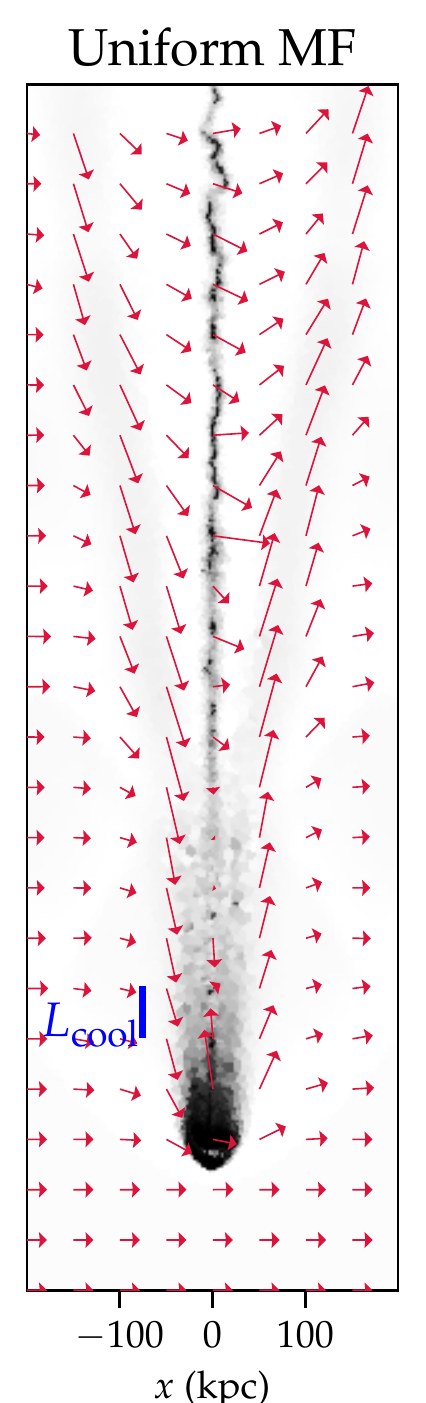}
        \includegraphics[width=0.275\textwidth]{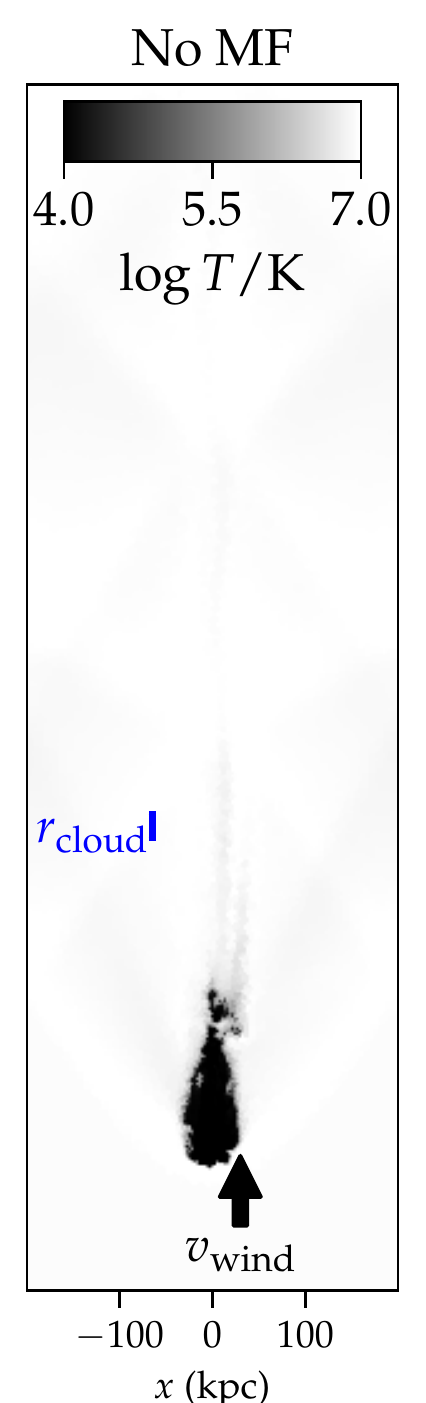}
    \caption{Slices of the temperature distribution overlaid with magnetic field (MF) vectors at four cloud crushing time scales. From left to right, we show simulations with a turbulent ICM wind magnetic field, an initially uniform magnetic field, and without any magnetic field interacting with a cold cloud with initial size $r_\textrm{cloud}$. In the latter case, Kelvin-Helmholtz instabilities disrupt the cold cloud while magnetic fields protect the cloud from disruption by creating magnetised filamentary structures that vary for different field topologies. The cold tail is considerably longer than the cosmic-ray electron cooling length (indicated in the left panel with effective $L_\textrm{cool}$).} 
    \label{fig:Temperature}
\end{figure*}

In our homogeneous-field simulation, magnetic tension of the draping layer back-reacts onto the flow and withstands alignment with the tail \new{at large distances from it (see Figure~\ref{fig:Temperature}). Interestingly, the field in the nearby tail region is highly ordered and almost uniformly assumes the theoretical maximum degree of polarisation of 0.75 for our assumed tail spectral index -- in contradiction to the observations.} In the turbulent-field simulations the magnetic tension is lower due to reversals of magnetic fields of different polarities, which are dynamically assembled in the magnetic draping layer. Thus, accretion of mixed warm gas to the cold phase in the wake aligns and stretches the field in the tail so that it \new{also} attains a high degree of polarisation, \new{approaching values of 0.6 in some tail regions with an overall patchier appearance which compares favourably to the observations.} Note that we do not model small-scale turbulence in the cold galactic cloud that is excited in star-forming regions and thus, our simulations overestimate the polarised intensity and fractional polarisation inside the clouds.

Figure~\ref{fig:Temperature} shows the temperature distribution of our simulations with the turbulent and the initially uniform magnetic field, and a purely hydrodynamical model without magnetic field. In the latter case, Kelvin-Helmholtz instabilities grow small perturbations at the wind-cloud interface and eventually disrupt the cloud. The magnetic draping layer protects the cloud from disruption and forms magnetised filamentary structures \cite{Dursi,Berlok}. The instabilities can act in the plane perpendicular to the initial orientation of the uniform magnetic field \cite{Dursi2007} via eddies that move field lines around while the turbulent field provides the best protection. The long radio tails should extend further into the downstream, which can be probed by low-frequency, high-sensitivity radio observations that map out older cosmic-ray electrons, which cool at a lower rate. Our simulations predict a much more extended cold tail (Fig. \ref{fig:Temperature}), which could be probed by future low-frequency radio and high-sensitivity HI measurements.

\section*{Discussion and Conclusions}

For the first time we analysed synchrotron intensity and polarisation of an extreme ram-pressure stripped galaxy based on deep observations and compared it to simulations. We discovered that the tail of the jellyfish galaxy emits radio emission with a steep spectrum ($\alpha\approx -2.0$), a high polarisation degree ($>50$\,\%), and a strong equipartition field ($>4.1\,\mu$G). In the disc, we estimate equipartition values between $6.7\,\mu$G and $7.1\,\mu$G. The high fractional polarisation and the magnetic field configuration imply a highly ordered magnetic field that is aligned with the extended tail. What is causing these long and ordered magnetic fields and what is the significance of the magnetic field on the formation of the tail and its on-going star formation?
 
To this end, we analysed magneto-hydrodynamic simulations of a galaxy-sized, cold gas cloud exposed to a hot, magnetised wind that the cloud feels as it moves through the ICM. We find the observed highly-ordered strong magnetic field far out in the tail to be in excellent agreement with our simulations, in particular with a turbulent ICM magnetic field (explaining our measured Faraday rotation, see SI, Fig.~8 and Sect.~7). Accretion of draped magnetised plasma from the hot wind that condenses onto the external layer of the tail, which is then adiabatically compressed and sheared by the velocity difference of the cold tail and the wind can explain the high degree of fractional polarisation and ordered magnetic field. There may be alternative mechanisms to produce the observed large-scale magnetic field in the tail and future studies of cosmological ram-pressure stripped galaxies are required to verify the suggested picture. 

These magnetised cold filaments in the jellyfish tail are then ``lit up'' by the synchrotron emission from cosmic ray electrons that are accelerated in supernova remnants in the interstellar medium and the star-forming tail. We find evidence in form of abundantly star-forming knots in the tail that show a spectral index flattening and reduced fractional polarisation. We conclude that these small, turbulent star-forming regions in the tail cannot disrupt the highly ordered magnetic field \new{(most probably due to our resolution element of $\sim15$\,kpc)}, which is aligned with the tail and illuminated by aged electrons with a steep spectral index. Such an ordered magnetic field may prevent heat and momentum exchange, and favor in-situ star formation in the tail. In this scenario, star formation in the tail should be a self-regulating process, achieving a balance with the magnetic field in a sort of feedback: a strong and widespread star-formation activity would disrupt the ordered magnetic field, which in turn is fundamental to allow the star-formation process to continue.

We demonstrate that turbulent wind magnetic fields are important for providing stability and shaping the ram-pressure stripped tails. Our simulations show that the polarised radio tail of the jellyfish galaxy should only be a small portion of a very extended magnetised and cold filamentary tail in the galaxy's wake, which might be detected with deeper observations. Higher-resolution observations are needed to characterise the role and connection of the magnetic field to the on-going star formation in the jellyfish tails.

\section*{Methods}\label{method}
{\bf Observations.} The total intensity and polarisation study in this paper is based on our new observation of JO206 with the Karl G. Jansky Very Large Array (JVLA) at 2.7\,GHz (project 18B-018, PI Poggianti), complemented for the study of the spectral index by the observation at 1.4\,GHz (project 17A-293, PI Poggianti) presented in \cite{Ramatsoku}.

The 2.7\,GHz observations were carried out during 11 nights between November 18th 2018 and December 15th 2018 with a total observing time of 50.5\,h and an on source time of 40\,h. All four polarisation parameters (RR, RL, LR, LL) were correlated. 
Details for each observation are summarised in Extended Table \ref{tab:observations}.
The observation parameters of the programme VLA/18B-018 can be found in Extended Table \ref{tab:parameter}.
We observed 3C48 for flux calibration, J2136+0041 as an intermittent phase calibrator, 3C138 as polarisation calibrator, and J2355+4950 for polarisation leakage calibration during each observing run.
%
%
We performed the data reduction using a combination of the Common Astronomy Image Software Application (CASA) and the Multichannel Image Reconstruction, Image Analysis and Display software package (MIRIAD).
Data were inspected using RFIgui \cite{Offringa}, the graphical front-end to aoflagger, to access the overall contamination with radio frequency interference (RFI). We found significant RFI contamination within the spectral windows 2, 3, 14, 15, and 16 for each observation. These were then flagged entirely. To remove spurious RFI signals in the rest of the observations we used aoflagger with a specifically tailored flagging strategy to our observations. In order to mitigate over-flagging due to roll-off effects at the edges of the sub-bands, we determined a preliminary bandpass using the unflagged flux calibrator. This bandpass was only used for flagging and not used in any of the later calibration steps.

We then followed the standard VLA cross-calibration scheme for each observing run separately. First, the data were corrected for antenna offsets where known. Flux, bandpass, and gain calibration as well as the removal of residual delays was performed using the flux calibrator source 3C48 with the model given in \cite{Perley13}. To correct for phase errors during the observation and for determining the polarisation leakage, which is introduced due to the rotation of the feeds with respect to the sky, we applied the previous solutions to our calibrator sources J2136+0041 and J2355+4950, respectively. The polarisation angle and cross-hand delays were then calibrated using the known polarised calibrator 3C138 \cite{Perley17}.

The standard synchrotron observables Stokes I (polynomial fit), Q, and U were then computed, while Stokes V was set to zero.
The coefficients for the polynomial fit as well as the fractional polarisation $p$ and polarisation angle $\chi$ for 3C138 are given in Extended Table \ref{tab:Polpara}.

Thereafter, the polarisation calibrator is used to solve for cross-hand delays (RL and LR). 
The total polarisation scale was solved by deriving the leakage solution from J2355+4950, denoted as leakage calibrator before.
We obtained an accurate polarisation position angle by calibrating the R-L phase using the polarisation calibrator, whose position angle is known from the model.
We transferred the flux scale from the primary calibrator to all secondary calibrators.
We applied all solutions to our target data sets, which were combined and imported into MIRIAD for self-calibration and imaging. 

Self-calibration was performed for each spectral window individually to mitigate imaging artefacts due to frequency dependent effects. We performed a pure imaging of the whole data set in total intensity to derive a mask for the sources. This mask was then used within the self-calibration and imaging in both cases to avoid including spurious sources.
Within the self-calibration process we used phase-only solutions, consecutively decreasing the solution interval and (u,v)-range included in the calibration process until a 30\,s interval was reached and all baselines were included. Each final individual image was convolved to the largest common beam and primary beam corrected. The final resulting image in total intensity was combined from the individual spectral window ones using an inverse square weighting based on their noise.  The final image was then inspected for not CLEANed-sources and a new mask derived to limit the CLEAN algorithm to regions of emission. The whole self-calibration procedure was iterated until no more sources in the final image showed sidelobe patterns. This ensures that the model for self-calibration is as complete as possible and fluxes of the target field are calibrated as accurately as possible.

For polarisation imaging we individually imaged the Stokes Q and U parameters. Each channel of 2\,MHz was imaged and cleaned (using the final mask from the total power imaging) separately in order to mitigate bandwidth depolarisation effects. We then corrected each image for primary beam attenuation. To include also the lowest frequencies to exceed the highest bandwidth and signal-to-noise ratio we convolved all our images to a circular beam of 15\,$^{\prime\prime}$.
The final 624 images in Stokes Q and U were used as an input to the Rotation Measure Synthesis algorithm.

Within this approach the data is transformed from frequency space into Faraday space via a Fast Fourier Transformation (FFT). This technique is mathematically similar to aperture synthesis imaging and known as Rotation Measure Synthesis \cite{Brentjens}. The instrumental parameters are therefore given by the frequency setup of the instrument and the instrumental function by the Rotation Measure Transfer Function (RMTF). With our setup we acquire a resolution in Faraday space of 223\,rad\,m$^{-2}$, a maximum observable scale of 458\,rad\,m$^{-2}$, and a maximum observable Faraday depth of 119816\,rad\,m$^{-2}$. We sampled the Faraday Q- and U-cubes between $-4096$\,rad\,m$^{-2}$ and $+4096$\,rad\,m$^{-2}$ with a step size of 16\,rad\,m$^{-2}$. We then calculated a polarised intensity cube, which is the absolute value of the complex polarised intensity $p=Q+iU$, where $Q$ and $U$ are the fluxes in the Faraday Q- and U-cubes.

Since a physical intensity can only be positive the polarised intensity is described by the complex conjugated
\begin{equation}
P = pp^* = \sqrt{Q^2 + U^2}    
\end{equation}
introducing a polarisation bias following a non-Gaussian statistic \new{\cite{1974ApJ...194..249W}. This bias is dependent on the distance from the pointing centre due to the primary beam correction. The bias in a polarised intensity map derived from RM-Synthesis cubes is higher than the theoretical one, the reason being the calculation of the polarised intensity from the maximum along the Faraday axis. This gives in case of only noise along this axis a higher chance of finding a higher value  \cite{2009A&A...503..409H}. The exact difference is therefore depending on the length of the sampled Faraday axis and the Faraday resolution of the observation. Hence,} we decided on an approximated approach to subtract the polarisation bias \cite{Adebahr}. 
We fitted a higher order polynomial with a parabola shape to the distance of a pixel from the pointing centre against the average value along the polarised intensity axis in the Faraday polarised intensity cube. Within this approach we only considered emission free areas, in the following denoted by the background. This parabola was then extended along the Faraday-axis to create a bias cube. Afterwards, we subtracted this cube from the polarised intensity one.

We derived the polarised intensity map and the Rotation Measure (RM) map (see SI, Fig. 8, right) by fitting a parabola along the Faraday-axis of the polarised intensity cube to any values exceeding 5\,$\sigma$. The peak value of the parabola gives the polarised flux density while the position of the peak in the Faraday spectrum gives the RM value. 
We derived the orientation of the electric field vectors with the polarisation angle $\psi$, which is given by
\begin{equation}
\psi = \frac{1}{2}\, \textrm{tan}^{-1} \frac{U}{Q},
\end{equation} 
with $Q$ and $U$ as Faraday rotation corrected Stokes parameter. Assuming that the RM and polarised intensity are originating from the same region, we de-rotated the electric field vector back to the intrinsic magnetic field one using
\begin{equation}
\psi_0 = \psi - \textrm{RM}\cdot\lambda ^2+90^\circ.
\end{equation}
While inspecting the background of the polarised intensity map we still found it to be higher than the standard deviation of $\sigma=2.0\,\mu$Jy/beam the background influencing the flux estimation. We corrected for this remaining bias (see SI, Sect. 5) and its result is shown in the left panel of Figure \ref{fig:Results2}.

In order to estimate the radio continuum spectral index of JO206 we combined the 2.7\,GHz image with a 1.4\,GHz made from a previous JVLA observation. The 1.4\,GHz data were taken with the goal of studying the neutral hydrogen gas properties of JO206 \cite{Ramatsoku}, and we refer the reader to that paper for further details on the 1.4\,GHz observation. Here we simply add that, for consistency with the analysis of the 2.7\,GHz data, for the present paper we calibrated and imaged the 1.4\,GHz continuum data using MIRIAD. Data reduction was carried out in a standard way. In particular, following cross-calibration and having discarded channels with HI emission, we iteratively imaged and self-calibrated the data (phase only, frequency-independent, and with a solution interval of 2 minutes) until the quality of the image converged. For this purpose, CLEANing was done using CLEAN masks made by carefully smoothing the image and applying detection thresholds that included real emission only. The continuum image used here was made using Briggs weighting with \it robust = 0 \rm resulting in a resolution of ($13.1\times 14.5)^{\prime\prime}$. This data is then convolved to the same angular resolution as for the 2.7\,GHz data resulting in a noise level of 40\,$\mu$Jy/beam.

An evaluation of the different angular scales and weightings of both frequency bands onto the properties of JO206 are discussed in Section 8 (see SI).
We find that this has no significant effect onto our scientific conclusions.
We therefore present the most complete images we were able to create to not miss any real diffuse emission. The presented values are derived from these images.

{\bf Simulations.} 
We intentionally choose very idealised simulations to focus on the dynamical interaction of a cold cloud with a a magnetized hot wind in the presence of radiative cooling rather than simulating a realistic galaxy with complex subgrid-scale models for feedback. Recent work shows that the warm phase (modelled here) is likely dominated by kinetic pressure over thermal pressure (see e.g., Fig. 9 of \cite{Gurvich2020} for an analysis of the FIRE2 simulations), implying that stripping may occur somewhat differently than in our simulations. In particular, while this internal structure and dynamics may have the potential to change the way the stripped gas mixes with the ICM, we believe that our 
scenario captures the main processes relevant to understand these new interesting observations. We fully acknowledge that a future high-resolution cosmological simulation of a galaxy orbiting in a galaxy cluster that follows the transport of the cosmic-ray electron spectrum will be able to scrutinize the scenario presented here.

Here, we provide additional details of the simulations. Initially, the cloud radius is $r_{\mathrm{cloud}}=\new{25} $~kpc, it has a temperature $T_{\mathrm{cloud}}=10^4$\,K, a number density of $n_{\mathrm{cloud}}=0.1\,\mathrm{cm}^{-3}$, and the hot ICM wind ($T_{\mathrm{wind}}=10^7$\,K) is in pressure equilibrium with the cloud so that the density contrast is $10^3$. The wind moves supersonically with a Mach number of $\mathcal{M}=2$, \new{yielding a wind velocity of $958~\mathrm{km~s}^{-1}$ which agrees with our observational estimates of $850~\mathrm{km~s}^{-1} < v \lesssim 1200~\mathrm{km~s}^{-1}$ within the systematic uncertainties (lower bound implies pure line-of-sight component and upper bound assumes 45$^\circ$ inclination of the galaxy velocity).}

The simulation code AREPO \cite{Springel,Pakmor13} solves the equations of ideal \mbox{(magneto-)}hydrodynamics on an unstructured Voronoi mesh \cite{Pakmor15} and employs the Powell scheme to ensure the divergence constraint of the magnetic field \cite{Powell}. If the mesh moves with the local fluid velocity, the scheme inherits the advantages of Lagrangian fluid methods that keep the mass per cell fixed. The code allows for inserting additional mesh-generating points that enable super-Lagrangian resolution capabilities and ensure the topological regularity of the individual Voronoi cells. To ensure a smooth transition in resolution between the larger cell size in the hot-diffuse wind and the cold-dense well-resolved cloud, we adopt a \emph{volume refinement criterion} ensuring that the volume of a cell does not exceed a factor of eight compared to the neighbouring cells. Adaptive time-steps are used to speed up the simulations. We use periodic boundaries in the $x$ and $z$ directions. The wind moves in the $y$-direction, and the wind properties are fixed in the \emph{injection region} near the lower $y$-boundary, as it is standard the procedure in simulations of a cloud interacting with a wind \cite{2015ApJ...805..158S,2017ApJ...834..144S}. We use a box size of $\{L_x , L_y , L_z\} = \{16 R_\text{cloud} , 384R_\text{cloud} ,16R_\text{cloud}\}$ and a target mass resolution of $8.6\times10^5~\mathrm{M}_\odot$ corresponding to 30 cells per cloud radius in the cloud.  

\new{The hydrodynamic simulations do not include any magnetic field.} For the simulations with a magnetic field included, we assume a plasma beta of $\beta=10$ everywhere in the simulations and adopt a tangled magnetic field inside the cloud \cite{McCourt}. We isolate the cloud's magnetic field from the wind, such that only tangential field lines are allowed at the cloud's surface \cite{2018MNRAS.481.2878E}. We iteratively remove the radial component and remove the divergence of the magnetic field, when creating the initial conditions. \new{We find that the magnetic topology inside the cloud is only of secondary importance for the formation of a magnetized tail; much more relevant is the magnetic topology of the hot wind: after being stripped from the galactic disk, the magnetic field is sheared and amplified in a narrow core region in the tail. By contrast, the wind magnetic field is draped around the galaxy and also amplified via velocity shear and condensation in the tail region where it dominates over the amplified galactic magnetic field \cite{Sparre2020}.} For the simulations with a uniform magnetic wind, we enforce the wind magnetic field of 0.6~$\mu$G in the injection region throughout the run time of the simulation.

For the simulations with a turbulent magnetic wind, we created a Gaussian field following a power spectrum of the form, $P_i(k) \propto k^2 \left|\tilde{B}_i(k)\right|^2$, with the absolute square of the Fourier transformation of the magnetic field component, $B_i$, being
\begin{align}
\left|\tilde{B}_i(k)\right|^2 = \begin{cases}
A,  & \text{ if } k< k_\text{inj},\\
A \left( \frac{k}{k_\text{inj}} \right)^{-11/3}, & \text{ if } k\geq k_\text{inj}.\\
\end{cases}
\end{align}
Here $k=2\pi/\sqrt{x^2+y^2+z^2}$ is the wave number, and the injection scale is $k_\text{inj}=1/(\sqrt{3}L_x)$. We hence have white noise on scales larger than the box-size, and on smaller scales, $k\geq k_\text{inj}$, we have Kolmogorov turbulence. The normalisation, $A$, of the power spectrum is chosen such that the magnetic field strength, $\sqrt{\langle \mathbfit{B}\rangle ^2}$, corresponds to $\beta=10$. The field is periodically initialised in the initial conditions and subsequently periodically read in the injection region.

Radiative cooling is modelled using the standard cooling function from the Illustris galaxy formation model \cite{2013MNRAS.436.3031V}. All gas cells are initialised with a solar metallicity.

\section*{Correspondence}
Correspondence and requests for materials should be addressed to AM (email: amueller@astro.rub.de). The JVLA 1.4\,GHz and 2.7\,GHz data will be available in the National Radio Astronomy Observatory (NRAO) archive (\url{https://science.nrao.edu/facilities/vla/archive/index}) and can be found via the project numbers stated in the methods.

\section*{Acknowledgements}
Based on observations collected at the European Organization for Astronomical Research in the Southern Hemisphere under ESO programme 196.B-0578. This project has received funding from the European Research Council (ERC) under the European Union's Horizon 2020 research and innovation programme (grant agreement No. 833824).
We acknowledge funding from the INAF main-stream funding programme (PI B. Vulcani) and from the agreement ASI-INAF n.2017-14-H.0.
This project has received funding from the European Research Council (ERC) under the European Union’s Horizon 2020 research and innovation programme (grant agreement no. 679627).
BV acknowledge the Italian PRIN-Miur 2017 (PI A. Cimatti).
CP acknowledges support by the European Research Council under ERC-CoG grant CRAGSMAN-646955.

\section*{Author contributions}
AM carried out the imaging and analysis of the 2.7\,GHz data as well as its interpretation, contributed to the comparison of the radio and optical data, and coordinated the research.
BP provided the MUSE data used in this paper and the measurements of the SFRD, contributed to the comparison of the radio and optical data and its interpretation.
AM, BP, and CP contributed to the text of the manuscript.
CP contributed to the various magnetic field estimates, MS run the simulations; MS and CP contributed to the analysis code, the synchrotron modelling and the interpretation of the simulations.
PS carried out the 1.4\,GHz data reduction and contributed to the interpretation of the results.
BA contributed to the 2.7\,GHz data reduction and its interpretation.
AI and MG carried out the analysis of the archival Chandra X-ray observation to asses the thermal properties of the cluster.
RJD contributed to the scientific discussion.
BV and AlMo contributed to the MUSE data acquisition and analysis, and the interpretation of the results.

\section*{Competing interests}
The authors declare that they have no competing interests.

\newpage

\newpage
\newpage
\newpage

\newpage
\newpage
\newpage

\renewcommand{\tablename}{Extended Table}

\begin{table*}[th!]
\centering
\begin{tabular}{rrrr}
\hline
Date & Time & Total time & Total down time \\
& & (min $\times$ 27 ants.) & (min $\times$ 27 ants.)\\
\hline
18Nov18 - 19Nov18 & 23:06:56 - 03:41:33 & 7414.7 & 0.0 \\
20Nov18  & 00:41:48 - 05:17:14 & 7436.7 & 0.0 \\
20Nov18 - 21Nov18 & 23:44:47 - 04:19:22 & 7413.7 & 0.0 \\
26Nov18 & 00:58:53 - 05:33:28 & 7413.8 & 271.8  \\
27Nov18 - 28Nov18 & 23:10:27 - 03:45:02 & 7413.7  \\
30Nov18 - 01Dec18 & 22:20:25 - 02:55:06 & 7416.4 & 0.0 \\
01Dec18 - 02Dec18 & 22:04:25 - 02:39:08 & 7417.4 & 549.4  \\
07Dec18 - 08Dec18 & 22:11:13 - 02:45:49 & 7414.2 & 480.5  \\
09Dec18 - 10Dec18 & 22:43:32 - 03:18:05 & 7412.9 & 274.6\\
11Dec18 - 12Dec18 & 22:41:10 - 03:15:44 & 7413.3 & 549.1  \\
14Dec18 - 15Dec18 & 22:29:23 - 03:03:57 & 7413.3 & 0.0 \\
\hline
\end{tabular}
\caption{Observations within programme VLA/18B-018. \label{tab:observations}}
\end{table*}
\begin{table}[th!]
\begin{tabular}{ll}
\hline
band & S (13\,cm)\\
configuration & C \\
polarisation products & Full \\
central frequency & 2.97\,GHz\\
total bandwidth & 2\,GHz \\
no. of antennas & 27 \\
no. of spectral windows (spw) & 16 \\
spw bandwidth & 128\,MHz \\
no. of channels per spw & 64 \\
total no. of channels & 1024 \\
channel bandwidth & 2 MHz\\
\hline
\end{tabular}
\caption{Observational parameter within the programme VLA/18B-018. \label{tab:parameter}}
\end{table}

\begin{table}[th!]
\centering
\begin{tabular}{rr}

\hline
Coefficients polynomial fit & \\
\hline
a$_0$ & 1.0088\\
a$_1$ & -0.4981\\
a$_2$ & -0.1552\\
a$_3$ & -0.0102\\
a$_4$ & 0.0223\\
\hline
Fractional polarisation & \\
$p$ / [\%] & 10.4\\
\hline
Polarisation angle & \\
$\chi$ / [deg] & -10\\
\hline
\end{tabular}
\caption{Polarisation parameter for 3C138. The values and a detailed descriptions of the coefficients of the polynomial fit can be found in \cite{Perley13} and the fractional polarisation as well as the polarisation angle in \cite{Perley17}.} \label{tab:Polpara}
\end{table}

\twocolumn[
\begin{@twocolumnfalse}
\section*{\centering \huge Supplementary information \vspace{1cm}}
\end{@twocolumnfalse}
]
\begin{figure*}[t]
\centering
\includegraphics[scale=0.82]{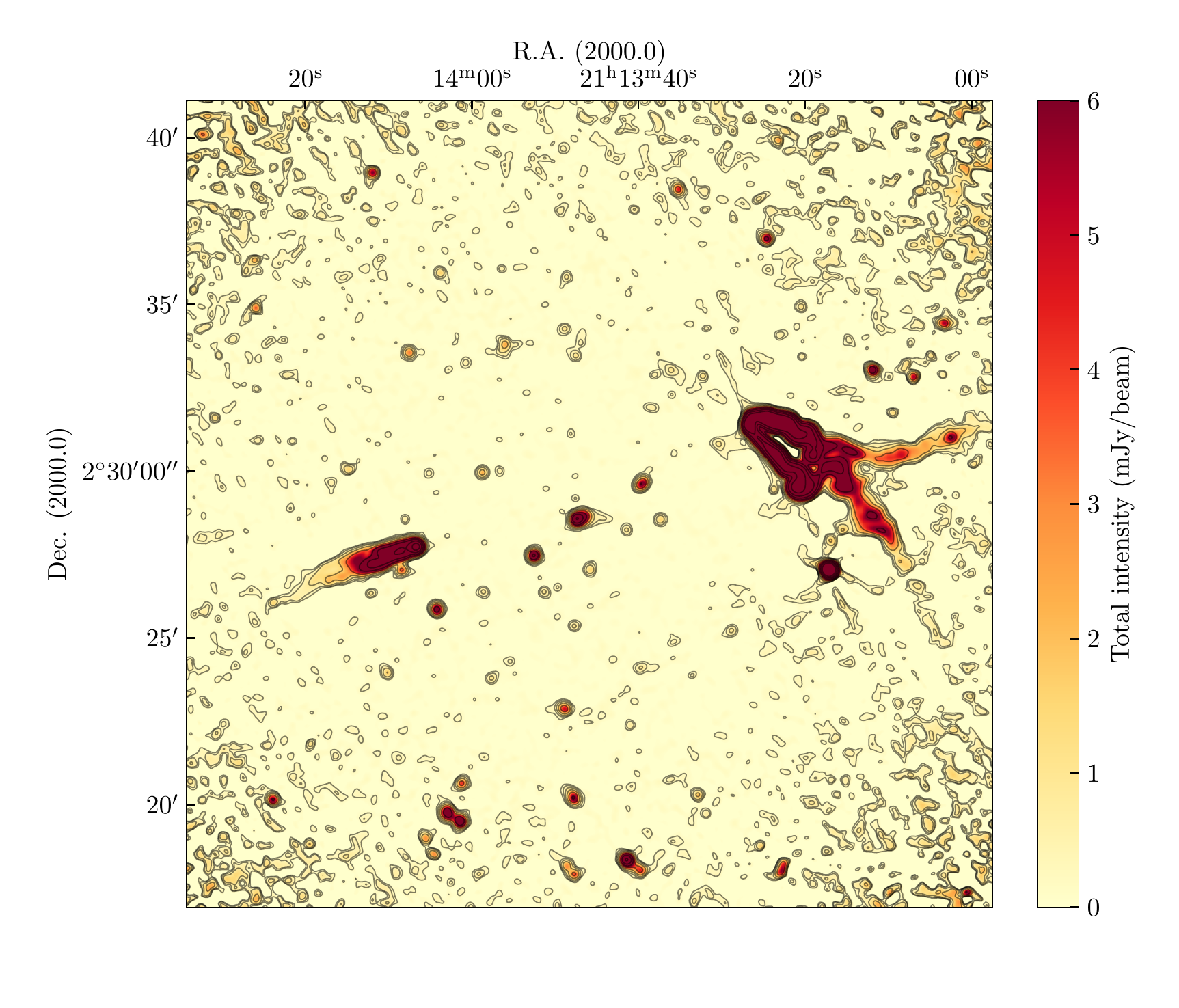}
\caption{Total intensity map at 2.7\,GHz of the field of view of the JVLA superimposed with contour levels of $\epsilon \times (1,\;2,\;4,\;8,\;16,\;32,\;64,\;128,\;256$) with $\epsilon = 21.5\,\mu$Jy/beam corresponding to 3\,$\sigma$.}
\label{fig:flux_fov}
\end{figure*}

\section{Overview of JO206}
\new{The jellyfish galaxy JO206 is undergoing extreme ram-pressure stripping in the low mass cluster IIZW108 with a projected distance of $\sim 300$\,kpc from the cluster centre \cite{Poggianti17SI,Poggianti17nSI}. It reaches a projected velocity of $850$\,km/s.
The 90\,kpc-long H$\alpha$-emitting tail of stripped gas of the galaxy JO206 shows significant ($\sim 0.5\,$M$_{\odot}$yr$^{-1}$) on-going, in-situ star formation \cite{Poggianti19SI}, while its disc star formation rate (4.8 $M_{\odot} \, yr^{-1}$) is normal for its stellar mass \cite{VulcaniSI}.
The gas metallicity in the disc is found to be supersolar, while it ranges from slightly subsolar to supersolar \cite{Poggianti18SI}.
For JO206 GASP has assembled a multiwavelength data set including optical integral-field spectroscopy with MUSE \cite{Poggianti17SI}, HI and radio continuum information with JVLA \cite{RamatsokuSI}, and molecular gas (CO) data with APEX \cite{MorettiSI} and ALMA (Moretti et al. in prep.).
To give an overview of the surrounding of JO206 the JVLA field of view is shown in Figure~\ref{fig:flux_fov}.}

\begin{figure*}[t]
    \begin{minipage}[t]{.5\textwidth}
        \centering
        \includegraphics[width=\textwidth]{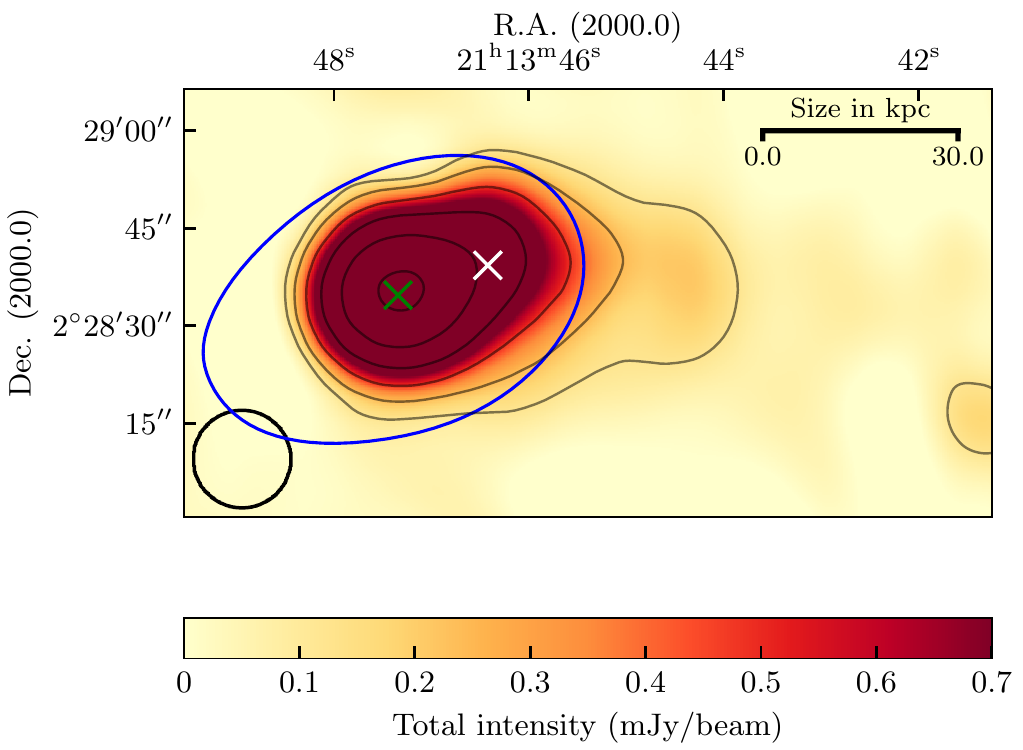}
    \end{minipage}  
    \hfill
    \begin{minipage}[t]{.5\textwidth}
        \centering
        \includegraphics[width=\textwidth]{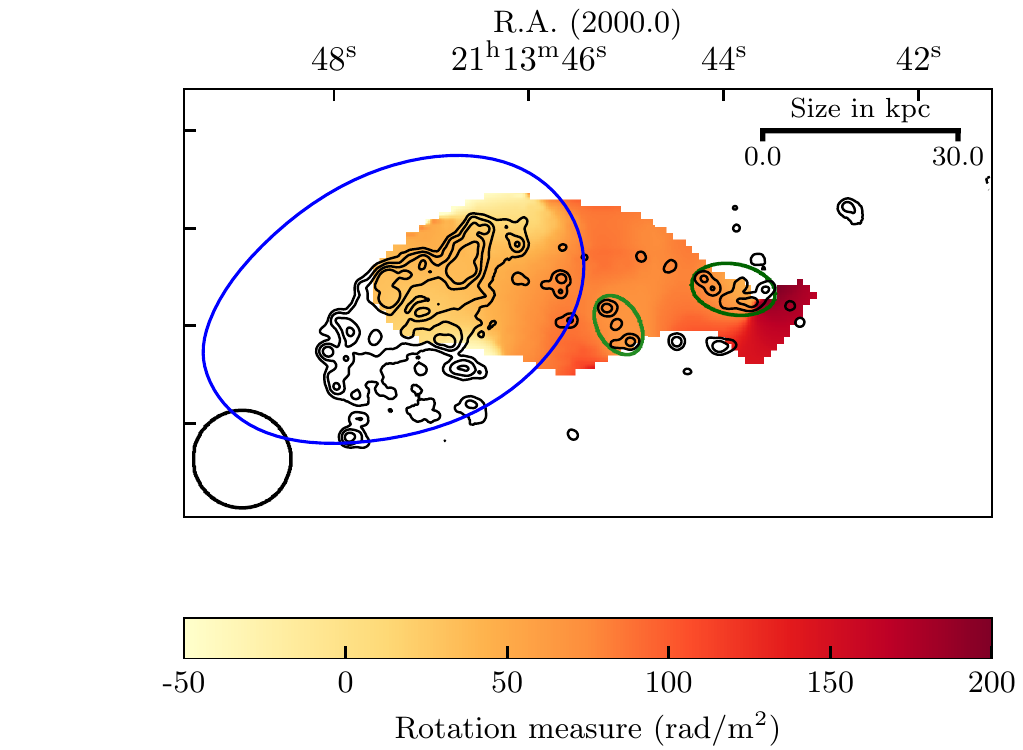}
    \end{minipage} 
    \caption{Additional observational results. The blue contour defines the optical stellar disc convolved with the 2.7\,GHz resolution. The white area contains no information. \textbf{Left}: Thermal corrected 1.4\,GHz total intensity map superimposed with contour levels of $\epsilon \times (1,\;2,\;4,\;8,\;16,\;32,\;64,\;128,\;256$) with $\epsilon = 120\,\mu$Jy/beam corresponding to 3\,$\sigma$. \textbf{Right}: RM map derived from the 2.7\,GHz data corrected for the galactic foreground superimposed with the contour levels of $\epsilon \times (1,\;3,\;9,\;27,\;81,\;283$) with $\epsilon = 1\cdot 10^{-19}\,$erg/s/cm$^2$ in H$\alpha$ and two star forming complexes are marked by a green and dark green ellipsoid.} \label{fig:Results4}
\end{figure*}
\section{Thermal contribution}\label{thermal}
The observed emission is not purely synchrotron but also affected by a thermal contribution mainly located in the star-forming regions.
A common strategy to correct for the thermal contribution to the radio total intensity is using dust corrected H$\alpha$ measurements \cite{DeegSI}.
The spatially resolved $\rm H\alpha$ emission is corrected for intrinsic dust extinction using the observed MUSE Balmer decrement (see Fig.~10 in \cite{Poggianti17SI}), assuming an intrinsic $\rm H\alpha$/$\rm H\beta$ ratio equal to 2.86 and adopting the \cite{CardelliSI} extinction law.

In Figure 1 (top left) and Figure \ref{fig:Results4} (left) we present our results corrected for thermal contribution at 2.7\,GHz- and 1.4\,GHz.
While the total intensity at 1.4\,GHz is only marginally affected by thermal electrons ($\sim 2\,\%$) we found the integrated total intensity at 2.7\,GHz (Fig. 1, top left) in the disc to be reduced by 12\,\% to ($3.5\pm0.2$)\,mJy while the tail emission is affected by about 30\,\% resulting in ($0.10\pm0.01$)\,mJy. Here, we take a typical calibration error of 5\,\% into account. 

We further identify the spectral index by the observed frequencies and integrated flux densities ($I_\nu \propto \nu^\alpha$) of the 2.7\,GHz and 1.4\,GHz observation.
After thermal correction, the values in the spectral index map (Fig. 1, top right) become even steeper resulting in an integrated spectral index $\alpha$ of $-0.74\pm0.09$ for the disc region and $-2.04 \pm 0.09$ for the tail ($-0.58 \pm 0.09$ and $-1.53\pm 0.09$ without thermal correction, respectively). A typical flux error of 5\,\% and the individual noise levels of the 2.7\,GHz- and 1.4\,GHz map are taken into account.

In the future a more reliable method should be investigated addressing the different opacity from radio and H$\alpha$ data. A correction for the line-of-sight integration is needed. 
Beside, an H$\alpha$ spectrum (which is used here) is less sensitive to diffuse emission than a photometric image (which is generally used). We have no information about the background noise (no detection corresponds to cut-off noise level in spectrum) in the H$\alpha$ map so that the statistics of the convolved H$\alpha$ map to the 2.7\,GHz resolution can be affected, therefore the resulting total flux image.

\section{Equipartition field strength}\label{limits}
The determined magnetic field strength based on the equipartition assumption [21] mainly suffers from the assumed pathlength and the spectral index limitations.
We cannot solve for spectral indices reaching values $>-0.54$, additionally, the magnetic field strength derived from spectral indices stepper than -1 become very uncertain, the assumption of $K_0=100$ is probably not valid for such steep electron spectra. Based on ref.~\cite{Beck05SI} we expect the tail values to be slightly overestimated. 
Second, getting a hold on the pathlength is very difficult since the geometry of our target is not known. JO206 is also not comparable to well-studied local spiral galaxies.
However, to restrict the magnetic field strength within the disc and tail we evaluated different pathlength for each region separately using the integrated thermal corrected total intensities, the spectral index (Sect. \ref{thermal}), and assuming a filling factor of one.

For the disc magnetic field we assume the scale height to be \new{between 1.1\,kpc and 1.4\,kpc} based on local spiral galaxies \cite{KrauseSI} and consider an inclination of 60\,$^\circ$ \cite{FranchettoSI}. \new{We, therefore, additionally assume that our target has a thick disc component and the synchrotron radiation being optically thin at the observed frequencies.
Based on the two different scale heights, we evaluate a pathlength of 2.2\,kpc and 2.8\,kpc, respectively resulting in a disc magnetic field strength of 7.1\,$\mu$G and 6.7\,$\mu$G, respectively.}
The tail geometry is assumed to be cylindrical. Based on the 2.7\,GHz radio continuum image, we measure a diameter of 28.5\,kpc as an upper limit for the pathlength. This provides us with a magnetic field strength of 4.1\,$\mu$G. Additionally, we assume a smaller diameter based on the higher resolved H$\alpha$ map of 20\,kpc resulting in a magnetic field strength of 4.4\,$\mu$G. However, we assume the tail material to be clumpy and therefore do not fulfill the constrain of a filling factor of one, which assumes the material to be continuous along the pathlength. Based on this parameter both values provide us with lower limits for the magnetic field strength in the tail. 

We therefore find the disc magnetic field lie between 6.5\,$\mu$G and 7.8\,$\mu$G and the tail magnetic field at a minimum of 4.1\,$\mu$G. In both cases the limitations for the spectral index are not evaluated. An appropriate method to calculate magnetic fields with steep spectral indices and complex geometries are beyond the scope of this work.

\section{Electron cooling timescales}
\label{sec:tau_cool}

The radio synchrotron emitting electrons of Lorentz factor $\gamma$ radiate at a
frequency
\begin{equation}
  \label{eq:nu_syn}
  \nu_\mathrm{syn} = \frac{3 e B \gamma^2}{2\pi m_e c},
\end{equation}
where $m_e$ is the electron mass, $c$ is the light speed, $e$ is the elementary change, $B$ is the root-mean-square value of the magnetic field. \new{This formula is} derived by replacing the synchrotron kernel with a Dirac delta function that is centered on this frequency. It reproduces the standard synchrotron result exactly for an electron spectral index of 3. Note that it differs from the critical synchrotron frequency by a factor of two. Synchrotron and inverse Compton aging of relativistic electrons occurs on a timescale
\begin{equation}
  \label{eq:tau_syn,ic}
  t_\mathrm{cool} = 
  \frac{6\pi m_e c}{\sigma_\mathrm{T} (B_\mathrm{cmb}^2+B^2) \gamma},
\end{equation}
where $B_\mathrm{cmb}\simeq3.2\mu\mathrm{G}$ is the equivalent field of the cosmic microwave background (cmb) energy density today and $\sigma_\mathrm{T}$ is the Thomson cross section. Combining both equations by eliminating the Lorentz factor $\gamma$ yields the cooling time of electrons that emit at frequency $\nu_\mathrm{syn}$,
\begin{eqnarray}
  \label{eq:taucool}
  t_\mathrm{cool} =
  \frac{\sqrt{54\pi m_e c\, e B \nu_\mathrm{syn}^{-1}}}
  {\sigma_\mathrm{T}\,(B_\mathrm{cmb}^2+B^2)}
  \lesssim1.3\times10^8 \,\mathrm{yr},
\end{eqnarray}
assuming an emission frequency of 2.7~\,GHz.  The cooling time $t_\mathrm{cool}$ is then bound from above and attains its maximum cooling time at $B=B_\mathrm{cmb,0}/\sqrt{3} \simeq 1.8\,\mu\mathrm{G}$, independent of the magnetic field.

The cosmic ray electron cooling time $t_\mathrm{cool}$ is shorter for fields weaker or stronger than $B= 1.8\,\mu\mathrm{G}$ due to the non-linear dependence of synchrotron cooling on the magnetic field, which yields two solutions that physically correspond to the dominant inverse Compton and synchrotron cooling regimes, respectively. Hence, the resulting cooling length of electrons depends on the magnetic field strength and is \new{$L_\mathrm{tail}(B)\simeq v_\mathrm{tail} t_\mathrm{cool}(B) \lesssim 27\,\mathrm{kpc}~(v_\mathrm{tail}/200\,\mathrm{km~s}^{-1})$, where $v_\mathrm{tail}$ is the velocity of the gas in the tail relative in the galaxy's rest frame in our simulation where we adopted a wind velocity of $958~\mathrm{km~s}^{-1}$. Scaling this result to a wind velocity of $1200~\mathrm{km~s}^{-1}$ yields $L_\mathrm{tail}(B)\lesssim 34\,\mathrm{kpc}~(v_\mathrm{tail}/250\,\mathrm{km~s}^{-1})$. In all cases, we assume a constant gas velocity along the tail.}

Assuming 1.\ that there is a negligible amount of electron re-acceleration along the tail, 2.\ a galaxy velocity of $850\,\mathrm{km\,s}^{-1}$ perpendicular to the line-of-sight (as inferred from the X-ray map, Fig.\ \ref{fig_X}) \new{which implies a three-dimensional wind velocity of $1200~\mathrm{km~s}^{-1}$ with an inclination of $45^\circ$} and 3.\ a constant magnetic field along the tail, we estimate the magnetic field to either be \new{$B\sim2.2\,\mu$G or $B\sim1.6\,\mu$G for the observed (projected) tail length of $L_\mathrm{tail}\simeq24$\,kpc of the radio synchrotron length.}

\section{Magnetic field strength in the front of the draping layer}\label{draping}
The prerequisite for a magnetic draping scenario is a galaxy moving super-Alfv\'{e}nically. Using a typical value for the cluster magnetic field of $B=1$\,$\mu$G, an electron density of $n_e=4.9\times10^{-4}$\,cm$^{-3}$, and a gas density of $\rho=9.5\times10^{-28}$\,g\,cm$^{-3}$, both derived from the Chandra X-ray data (see Sect.\ \ref{X-ray}), we determine an Alfv\'{e}n velocity of $v_\mathrm{A} = B/\sqrt{4\pi\rho} = 90$\,km\,s$^{-1}$. The morphology of the galaxy suggests a considerable in-sky velocity component transverse to the line-of-sight, $v_\perp$. We therefore assume $v_\perp$ to be of order of the line-of-side velocity component, $v_\parallel \approx850$\,km\,s$^{-1}$ \cite{Poggianti17nSI} so that we obtain the total velocity of the galaxy as 1200\,km\,s$^{-1}$. We furthermore assume that the galaxy moves supersonically through the ICM with a sonic Mach number of at least 1.3 (see Sect. \ref{X-ray}) so that we obtain a post-shock velocity of $v_\mathrm{post}\approx \new{830}$\,km\,s$^{-1}$ after applying the Rankine-Hugoniot jump conditions. This yields an Alfv\'{e}nic Mach number for the galaxies motion in the intracluster medium of \new{$M_\textrm{A}=v_\mathrm{post}/v_\mathrm{A} \approx 9$}. 
This further allows us to calculate the magnetic field strength at the stagnation point of the draping layer,
\begin{equation}
\label{eq:Bmax}
B_\textrm{max}=\sqrt{8\pi \alpha \rho v_\mathrm{post}^2}=\new{18}\,\mu \mathrm{G},  
\end{equation}
where $\alpha=2$ is a geometric factor inferred from magneto-hydrodynamic simulations of magnetic draping \cite{DursiSI}.

\section{Bias evaluation}\label{bias}
While computing the polarised intensity data cube a bias is induced, \new{which generally increases the mean of the background (for more details see Methods)}. Such bias can affect significantly the low surface brightness of targets, here, in particular, the jellyfish tail emission.
We already subtracted for such induced bias based on a least-square-fit to the mean of the background (see Methods). 
However, we recognize that the mean of the background in the surroundings of JO206 (pointing centre) is still higher than the signal-to-noise ratio and therefore expected to be affected by a remaining bias. 
We cannot rule out that we do not observe additional emission originating from the jellyfish galaxy or the ICM but here we simply assume that the remaining background level is only due to the induced bias.
Assuming that the background level is only due to a bias provides us with lower limits for the polarised emission and fractional polarisation and a maximum deviation to the values that can be derived from the non-corrected data.

We assume that the mean of the background in the surroundings of the jellyfish galaxy is a measure for the bias since no significant gradient is found in there.
We then subtracted the data by this bias ensuring that no negative values occur in the data. 
We therefore find the remaining bias to be $\sim 5$\,$\mu$Jy/beam. 
The polarised intensity and fractional polarisation map corrected for the remaining bias are shown in Figure 2 left and right, respectively. 
The integrated polarised intensity is found to be ($0.136\pm 0.007)\,$mJy from which ($0.083\pm 0.004)\,$mJy originates from the disc and ($0.054\pm 0.003)\,$mJy from the tail. Without this additional bias correction we measure a total polarised intensity of ($0.168\pm 0.008)\,$mJy, ($0.098\pm 0.005)\,$mJy originating from the disc and ($0.070\pm 0.004)\,$mJy  from the tail. The background therefore affects the polarised emission by about 18\,\% at maximum. The errors are again based on the 5\,\% calibration uncertainty.
We estimate the fractional polarisation based on the integrated thermal corrected total intensity and polarisation reaching ($2\pm 6$)\,\% in the disc and ($54\pm 10$)\,\% in the tail (Fig. 2, right). We find maxima $>60$\,\% where the total intensity drops very fast, along the western edge of the disc.  Without this additional bias correction we find a fractional polarisation of ($3\pm6$)\,\% in the disc region and ($70\pm10$)\,\% in the tail. 
By reducing the background assuming it to be caused by an induced bias the fractional polarisation is reduced by $\sim 20$\,\%, but qualitatively our conclusions remain the same.

\section{Rotation measure}
\label{sec:RM}

We also computed the rotation measure (RM) map (Fig.~\ref{fig:Results4}, right panel), which is an observable for the frequency of rotation of the electric field vector along the line-of-sight due to Faraday rotation. Positive values represent a magnetic field that is pointing towards us while negative ones describe the opposite. The RM map is corrected for the galactic foreground which, based on \cite{OppermannSI}, reaches 5.8\,rad\,m$^{-2}$ at the galaxy position in the sky. The observed RM values oscillate between -50\,rad\,m$^{-2}$ and +50\,rad\,m$^{-2}$ towards the disc while we find significantly higher values towards the tail of $\textrm{RM}\approx100$\,rad\,m$^{-2}$. The extreme values of 200\,rad\,m$^{-2}$ in the most distant part of the tail are very uncertain due to the low polarised signal-to-noise ratio. 

We postpone a detailed RM modelling of the tail to future work but we still give an estimate of the RM contribution of the turbulent and magnetised ICM (in the cgs system of units) via:
\begin{equation}
\textrm{RM} = \frac{e^3}{2 \pi m_\mathrm{e}^2 c^4} \int \limits_{r_1}^{r_2}  n_\textrm{corr}(r) B(r) \frac{r\,\textrm{d}r}{\sqrt{r^2-r_1^2}},  
\end{equation}
where $r$ is the cluster-centred radius, $r_1\approx280$\,kpc is the projected galaxy position (assuming that the galaxy is located in a plane perpendicular to the line-of-sight that cuts through the cluster centre), and $r_2\approx2400$\,kpc is our assumed outer cluster boundary (which corresponds to $3 R_{500}$). We verified that the resulting RM values are insensitive to the exact choice of this parameter. The ICM magnetic field $B$ is modeled by
\begin{equation}
B(r) = B_0 \, \left[\frac{n_\textrm{corr}(r)}{n_0}\right]^{\alpha},     
\end{equation}
where $B_0=1\,\mu$G is the central value of the root-mean-square magnetic field, $n_0$ is the central electron density, and $\alpha=0.7$. The electron density profile derived from X-ray observations (Sect. \ref{X-ray}) is only measured out to a radial distance of 585\,kpc. Extrapolating the density measured at these small radii towards the outer cluster regions (using the shallow density slope) would over-estimate the density and hence the RM signal. We therefore correct the electron density profile at larger radii and obtain
\begin{equation}
n_\textrm{corr}(r)=n_\mathrm{e}(r)\, \Bigg[1+\bigg(\frac{r}{1.5\,r_\textrm{out}}\bigg)^{\delta}\Bigg]^{-3(1-\beta)/\delta},
\end{equation}
where $n_e(r)$ is the measured electron density profile of the cluster (Sect. \ref{X-ray}), $\delta=3$, $\beta=0.44$ (Sect. \ref{X-ray}), and $r_\textrm{out}=585$\,kpc as the outer radius where we still were able to measure the electron density. This results in a steeper slope for large radii, which is expected based on theoretical and observational findings of deeper X-ray observations (e.g., \cite{LauSI, MorandiSI}). We estimate an RM contribution of the ICM of 65\,rad\,m$^{-2}$, which matches our observed RM values and is also consistent with our simulation choice of a turbulent ICM wind magnetic field. Only a small portion of the observed RM is originating from the galaxy.

\section{\new{High resolution imaging}}
\begin{figure*}[t]
    \begin{minipage}[t]{.5\textwidth}
        \centering
        \includegraphics[width=\textwidth]{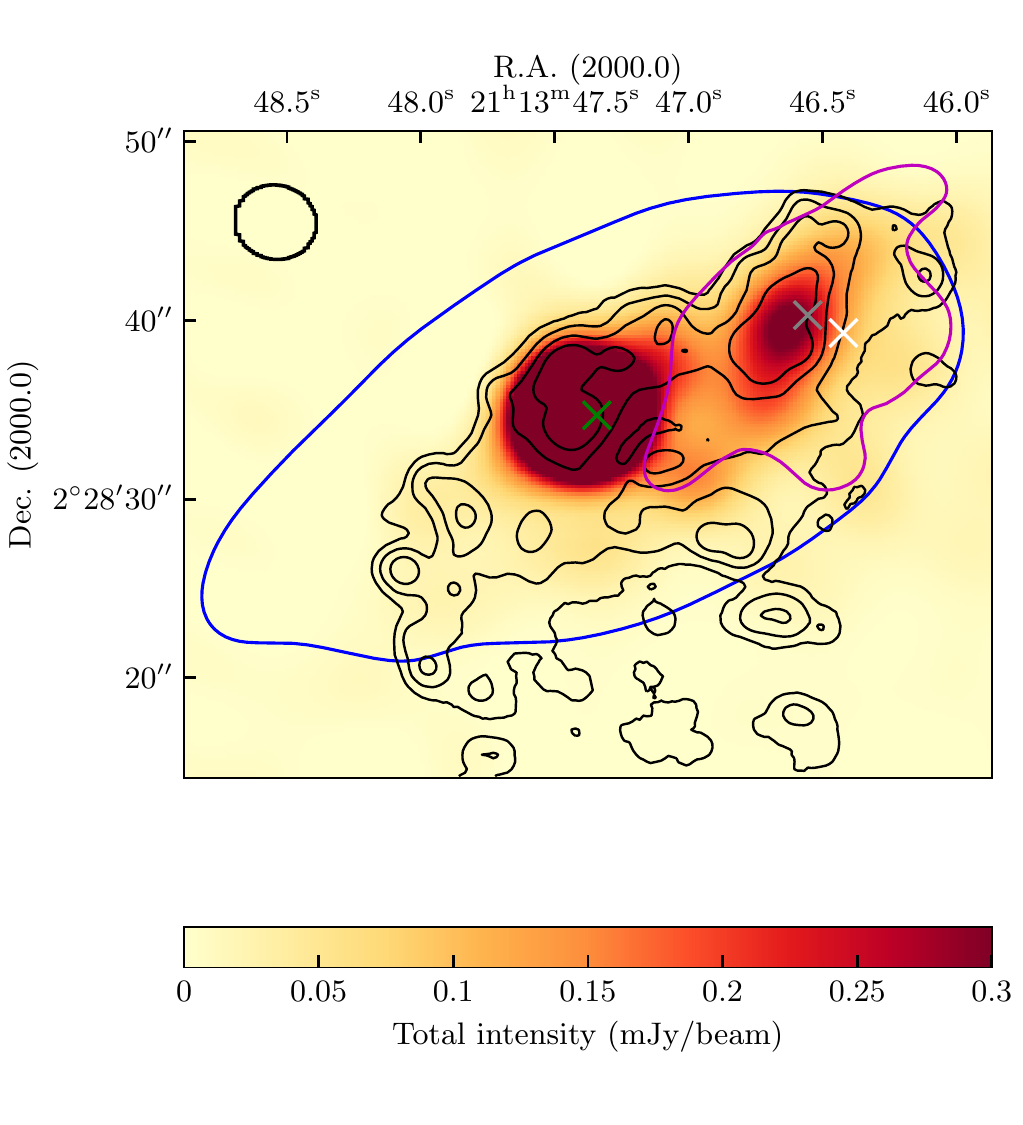}
    \end{minipage}  
    \hfill
    \begin{minipage}[t]{.5\textwidth}
        \centering
        \includegraphics[width=\textwidth]{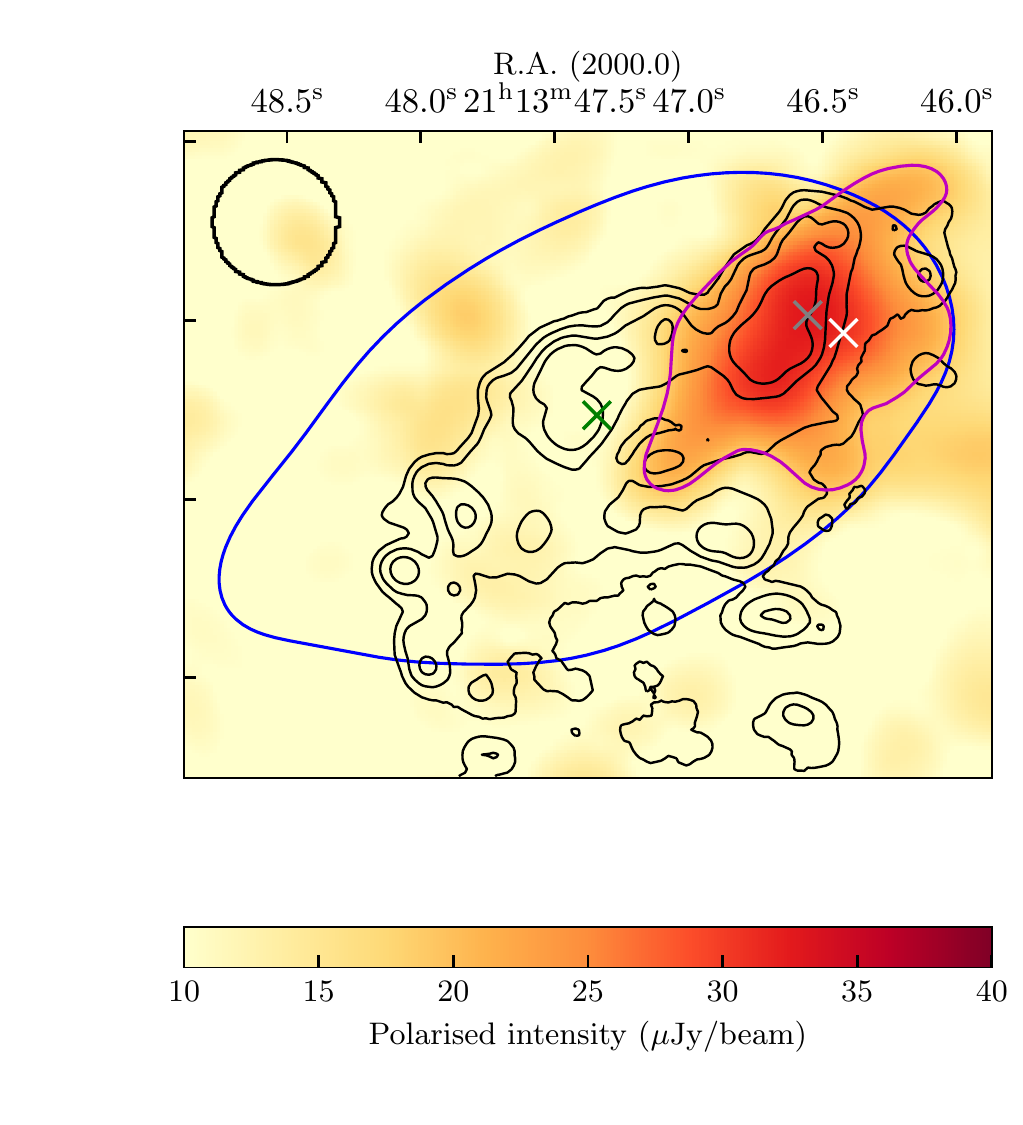}
    \end{minipage} 
    \caption{\new{High resolution total and polarised intensity maps. The stellar disc is defined by a blue contour convolved to the resolution of the corresponding image. Superimposed we show the black contour levels of $\epsilon \times (1,\;3,\;9,\;27,\;81,\;283$) with $\epsilon = 1\cdot 10^{-19}\,$erg/s/cm$^2$ in H$\alpha$ and a magenta contour level of 4.5 $\times$ the noise level of the polarised intensity. The peak emission of the total and polarised intensities found in the low resolution images for orientation are shown in green and white, respectively. With a gray cross the peak of the high resolution polarised intensity map is marked. \textbf{Left}: Total intensity map at a resolution of $4.59^{\prime\prime}\times4.15^{\prime\prime}$. \textbf{Right}: Polarised intensity map at a resolution of $7.0^{\prime\prime}\times7.0^{\prime\prime}$.}} \label{fig:highres}
\end{figure*}

\new{To evaluate the origin of the disc flux and of the spatial offset between the peak emission in total and polarised intensity we generated images at the highest resolution that still results in a significant detection (at least 3\,$\sigma$ in total intensity and 4.5\,$\sigma$ in polarised intensity; Fig. \ref{fig:highres}). We settled on a resolution of $4.59^{\prime\prime}\times4.15^{\prime\prime}$ in total intensity and $7.0^{\prime\prime}\times7.0^{\prime\prime}$ in polarised intensity, resulting in a noise level of 4.1\,$\mu$Jy and 2.1\,$\mu$Jy, respectively. Here, we focus on the disc of JO206 since no emission is detected in the tail at this higher resolution.}

\new{Two main components contribute to the overall total-intensity emission in the disc: the central source, which is most likely AGN dominated, and the strongest star-forming regions west of the AGN. At high resolution, the integrated total-intensity flux density of the disc is ($3.4\pm0.2$)\,mJy, which is the same as in the low resolution image within the uncertainties. The high-resolution bright, central source is co-located with the total-intensity peak at low resolution. Therefore, we conclude that the disc total-intensity emission detected at low resolution ($15^{\prime\prime}$) includes a significant contribution from both the central source and the bright star-forming regions.}

\new{At high resolution, the integrated polarised-intensity flux density is ($0.062\pm0.003$\,mJy), which is about 0.02\,mJy less than in the low resolution image. The polarised intensity peak (white cross in Fig. \ref{fig:highres}) of the low resolution image is slightly shifted west (about $2^{\prime\prime}$) from the center of the bright star-forming region in the high-resolution total intensity map (Fig. \ref{fig:highres}, panel a). This is most probably due to the fact that, at low resolution, emission from the tail is superimposed with that from the disc. The peak in polarised emission at high resolution (Fig. \ref{fig:highres}, panel b), where the tail emission contributes less, is more clearly co-located with the star-forming region (see cyan cross in Fig. \ref{fig:highres}).} 

\new{The higher resolution image of the polarised intensity still does not show any emission in the northern, eastern, and southern part of the disc. Both the high and low resolution image most probably suffer from beam depolarisation, indicating a high turbulent component in the northern, eastern, and southern disc region. 
Our resolution element encompasses many turbulent cells all along the "impacting edges". There, the ram-pressure stripping directly impacts the ISM and therefore the turbulence is likely high. Another possibility would be a toroidal disc field in the galaxy, where the magnetic field vectors change direction quite rapidly at the edges of the disc. This would cause an overlap of vectors of different direction both at high- and low resolution, and therefore depolarisation (see \cite{2007A&A...471...93WSI} for examples in the Virgo cluster). Deep high resolution observations are needed to examine this issue in more detail.}

\section{Common spatial resolution}\label{uvcut}
Generally, an interferometer is a spatial filter, where the recovered scales in an image are dependent on the baseline length and observed frequencies. 
We verify for the difference in recovered spatial scales by imaging the 2.7\,GHz and 1.4\,GHz observations by using the same Briggs weighting with \it robust = -2 \rm. In addition, we determine the minimum and maximum common $(u,v)$-distances and imaged the overlapping $(u,v)$-coverage for both frequencies individually. For the 2.7\,GHz observation we found no significant difference for the integrated total intensity while the 1.4\,GHz flux shows a decrease by 0.1\,mJy in the tail region due to the (now) missing shortest baselines.
Short baselines are necessary to reconstruct the diffuse emission of a source and we therefore conclude that the 2.7\,GHz data is suffering from missing short spacings.

In the following, we expect the 1.4\,GHz data to recover the complete integrated flux of JO206 while we measure lower limits for the fluxes at 2.7\,GHz. 
Based on the reduced integrated total flux at 1.4\,GHz the spectral index and magnetic field strength in the tail is affected while the integrated disc values stay the same.
We find a spectral index of $-1.76 \pm 0.09$ in the tail resulting in a magnetic field strength of $3.4\,\mu$G and $3.1\,\mu$G based on a pathlength of 20\,kpc and 28.5\,kpc, respectively. We therefore find a slightly flatter spectral index still indicating a significantly aging of the electrons propagating from the jellyfish disc into the tail. The tail magnetic field provides us with a lower limit of $3.1\,\mu$G, which is still, within the systematic uncertainties, in agreement with the high-field solutions.

\section{Establishing our simulation model}

\new{ Here, we provide energy and timescale arguments in support of our model to correctly capture the physics essentials of ram-pressure stripping of galaxies. While the inclusion of a disky galaxy morphology \cite{Pfrommer10SI,RuszkowskiSI}, self-gravity \cite{Aung2019SI} and gravity due to the stars likely changes quantitative aspects of the instability criterion \cite{Mandelker2019SI,Mandelker2020SI}, the inclusion of magneto-hydrodynamics \cite{BerlokSI} and specifically radiative cooling \cite{GronkeSI,Gronke2SI,Gronke3SI,Sparre2019SI,Sparre2020SI,Li2020SI} are believed to be the most important physics ingredients responsible for the formation and morphology of the ram-pressure stripped tails. The physics of magnetic draping of the ICM magnetic field at the galaxy-wind interface also is not qualitatively altered by considering a disky galaxy morphology \cite{DursiSI,Pfrommer10SI}.}

\begin{figure*}[t]
    \begin{minipage}[t]{.5\textwidth}
        \centering
        \includegraphics[width=\textwidth]{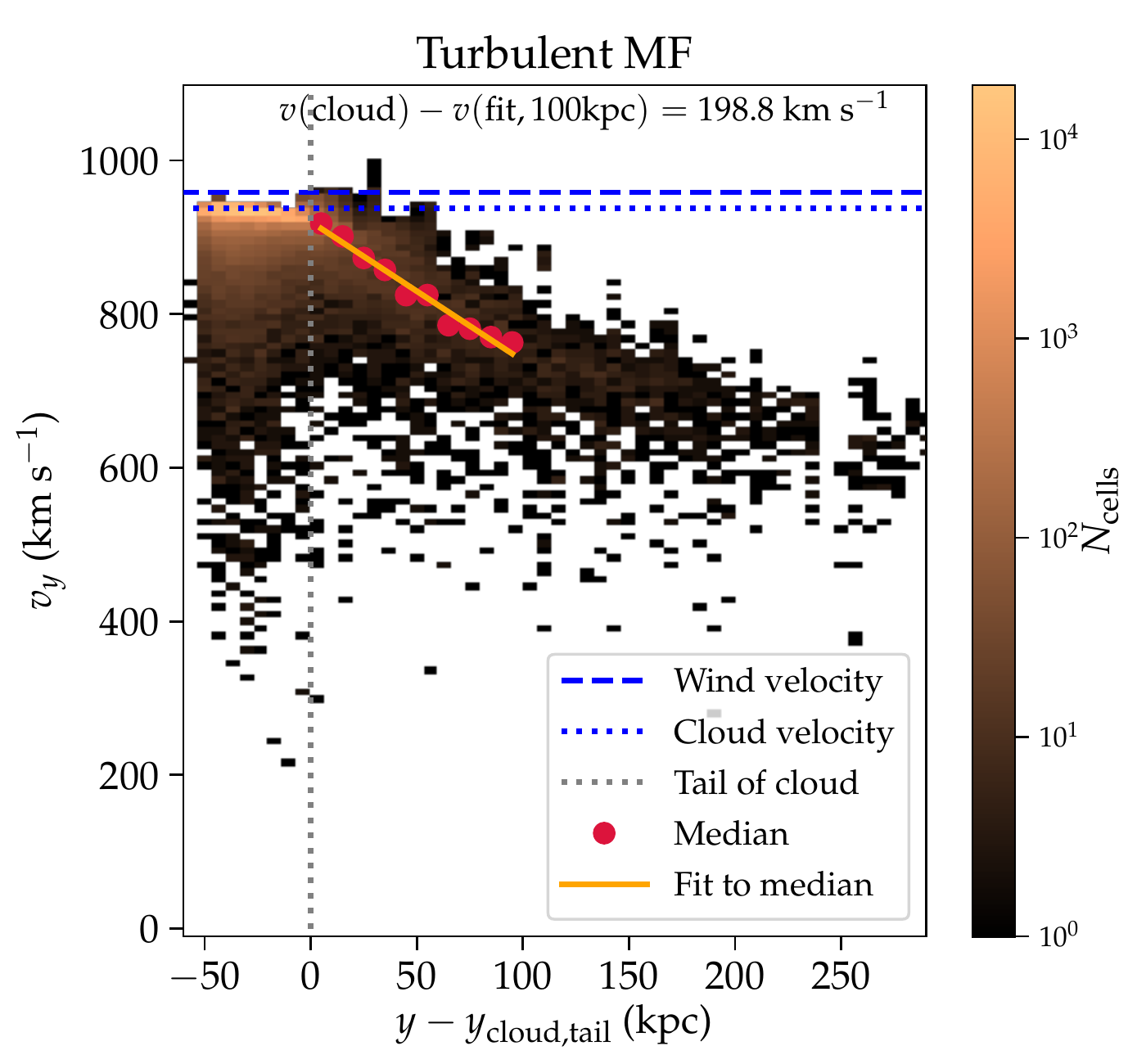}
    \end{minipage}  
    \hfill
    \begin{minipage}[t]{.5\textwidth}
        \centering
        \includegraphics[width=\textwidth]{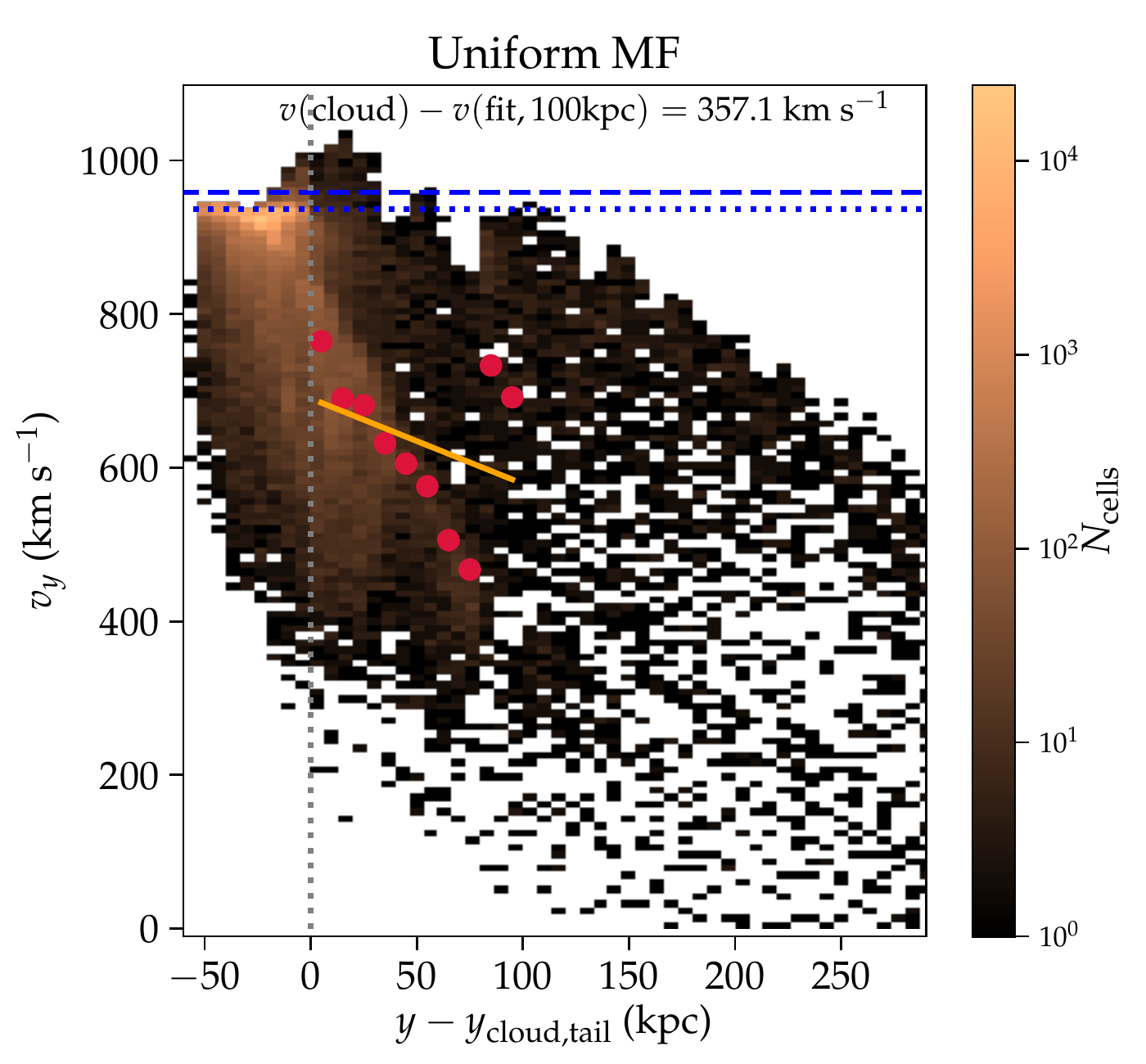}
    \end{minipage} 
        \caption{\new{Phase space of the distance and velocity along the tail for the initially turbulent (left) and uniform (right) wind magnetic field. We show the velocity of gas cells, which have been part of the initial cloud and got ram-pressure stripped, as a function of the distance downstream of the cloud tail at four cloud crushing timescales. We fit a linear relation and use this to determine the relative velocity difference between the cloud and the downstream gas.}}
        \label{VelProfileJellyfish_vy_25kpcM2.0_Turbulent_MF_M2.0_064}
\end{figure*}

\new{To decide whether the shear at the interface of the ram-pressure stripped ISM and the hot ICM wind or the rotational motion of a disk galaxy dominates, we have to compare the timescales at a given length scale, i.e., we need to compare the velocities. We find a velocity shear along the tail of about $750-800~\mathrm{km~s}^{-1}$ (in the cluster reference frame, see Fig.~4 in the main paper and Fig.~\ref{VelProfileJellyfish_vy_25kpcM2.0_Turbulent_MF_M2.0_064}) which dominates over rotation velocities of around 220 km/s so that the kinetic shear energy dominates over the rotation energy by a factor of $12-13$, justifying our neglect of the effect of rotation to first order. Using Occam's razor, energy arguments and very recent literature on the cloud-wind interaction, we believe that we included the most relevant physics and the success of explaining the main aspects of the simulations (magnetic alignment and polarisation fraction) supports our transparent, minimum physics approach.}

\begin{figure}[t]
        \includegraphics[width=0.5\textwidth]{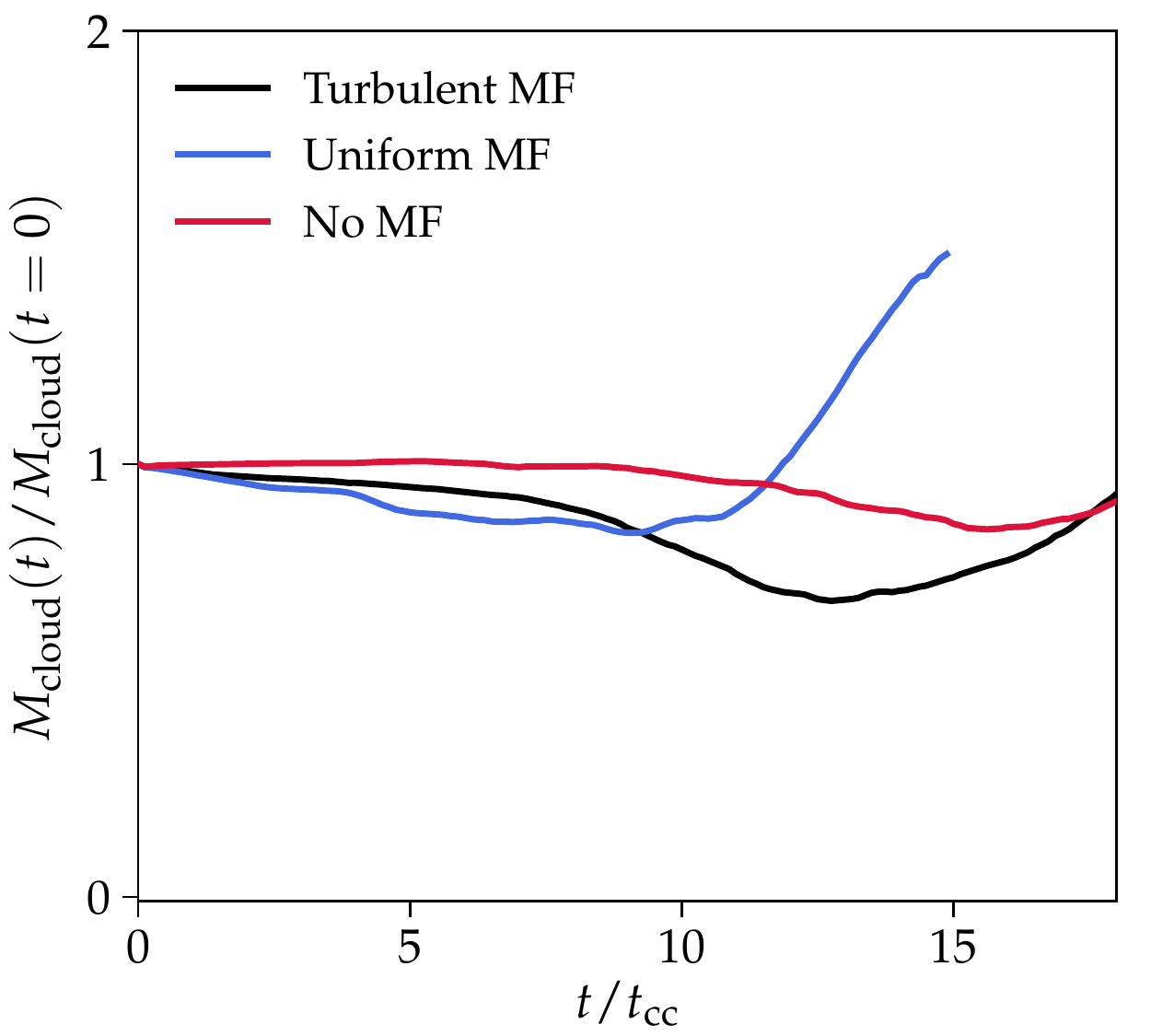}
        \caption{\new{The evolution of the mass in the dense phase with a density larger than a third of the initial cloud density. All simulations are in the growth regime because the clouds experience at least one episode of continuous cloud growth.}}
        \label{FigX122_CloudGrowth_NatureReport_3D}
\end{figure}

\new{Cloud-crushing simulations are often used to simulate how a hot wind destroys a cold cloud \cite{2017ApJ...834..144SSI}. It has, however, recently been established that for sufficiently large clouds, there is a net accretion of hot gas to the cold--dense phase, such that the mass in the cold--dense phase grows in time \cite{GronkeSI,Li2020SI,Sparre2020SI}. In our simulations we see such a growth occurring in the tail of the clouds -- see Fig.~6 of the main paper -- such that the clouds develop a structure comparable to a jellyfish galaxy. To quantitatively establish that growth occurs in the simulations, we show the mass of the dense phase with $n\geq n_\text{cloud}/3$ as a function of the cloud crushing time-scale (Eqn.~1 in the main paper) in Fig.~\ref{FigX122_CloudGrowth_NatureReport_3D}. We plot the evolution either until $18t_\text{cc}$ or until gas in the dense phase starts leaving the simulation box. For the simulations with a uniform magnetic field in the wind (\emph{Uniform MF}) the gas starts growing earlier than the other simulations. In this simulation dense gas is accelerated far downstream at around $\gtrsim 15t_\text{cc}$ so that dense gas is leaving the simulation box. Consequently, we do not show this simulation beyond $15t_\text{cc}$. All simulations show a net growth at late times, independent of the magnetic field used, and are thus in the growth regime.}

\section{Modelling of cosmic ray electrons}


\new{To generate a mock synchrotron emission map we need an estimate of the cosmic ray electron cooling length in the tail of the cloud. First, we determine the Lagrangian volume that we use to compute the cooling length. At four cloud crushing timescales ($t_\text{cc}$, see Eqn.~1 in the main part), where we create our synchrotron maps, we select gas cells originating from the cloud at the start of the simulations. We identify the $y$-coordinate of the head of the cloud as the 0.1 percentile of $y$ for this gas reservoir. The tail of the cloud starts then at a distance of $2R_\text{cloud}$ downstream of the head. In Fig.~\ref{VelProfileJellyfish_vy_25kpcM2.0_Turbulent_MF_M2.0_064} we show $v_y$ as a function of $y$ for both simulations with the turbulent and uniform magnetic field, and we have marked the location of the cloud's head and tail. From 0 to 100 kpc downstream of the tail, we calculate the median $v_y$ in 10 bins equally spaced in $y$. We fit a linear function to these median values, and by using the fitted function we determine that stripped gas at 100 kpc downstream of the cloud has a typical velocity of $200$~km~s$^{-1}$ and 360~km~s$^{-1}$ relative to the head of the cloud (in the turbulent and homogeneous field simulations, respectively).} 

\new{To estimate the cooling time of the downstream gas we use Eqn.~\ref{eq:tau_syn,ic}. Between 0 and 100 kpc downstream of the cloud's tail, we measure the median magnetic field of gas originally belonging to the cold cloud, and we obtain a value of around $1~\mu$G for the turbulent magnetic field simulation. Substituting this value into Eqn.~\ref{eq:nu_syn} of the SI and multiplying with the downstream velocity of $200$ km~s$^{-1}$, we obtain a cooling length of 25~kpc. For the simulations with a uniform magnetic field we measure (at $4t_\text{cc}$) a magnetic field and the relative velocity to be $2.1~\mu$G and 360 km~s$^{-1}$, respectively. We show this in Fig.~\ref{VelProfileJellyfish_vy_25kpcM2.0_Turbulent_MF_M2.0_064} (lower panel) and we get a cooling length of 50~kpc. Our choice of simulation parameters yields a magnetic field strength at the stagnation point of the draping layer of $B_\mathrm{max}=3~\mu$G, which falls short of the observational inferences by a factor of six (see Eqn.~\ref{eq:Bmax}). If the magnetic field strength of the tail would be solely determined by the ram pressure, we would have to scale our field strengths by the same factor to obtain a realistic magnetic field of the tail of JO206. However, future work is needed to establish the non-linear map from $B_\mathrm{max}$ to the tail magnetic field strength and to establish the relative roles of velocity shear, compression due to gas condensation, and potentially a small-scale dynamo in setting the saturated magnetic strength.}

\new{We postpone a detailed transport of the cosmic-ray electron spectrum \cite{WinnerSI} that takes into account the full distribution of the velocity and magnetic fields along the Lagrangian trajectories to future work. This will then be able to model excursions of pockets of low and high velocity values that may modify the radio-emitting length
(note that this shows a broad distribution, see Fig.~\ref{VelProfileJellyfish_vy_25kpcM2.0_Turbulent_MF_M2.0_064}). In principle the cosmic ray electrons in the tail may also experience gentle Fermi-II re-acceleration \cite{deGasperin2017SI}, which may partially counteract synchrotron and inverse Compton cooling and thus increase the length of the radio-emitting tail. We refrain from adding this complication because the data does not convincingly require this degree of freedom.}

\section{X-ray analysis}\label{X-ray}
\begin{figure*} 
\includegraphics[width=0.49\textwidth]{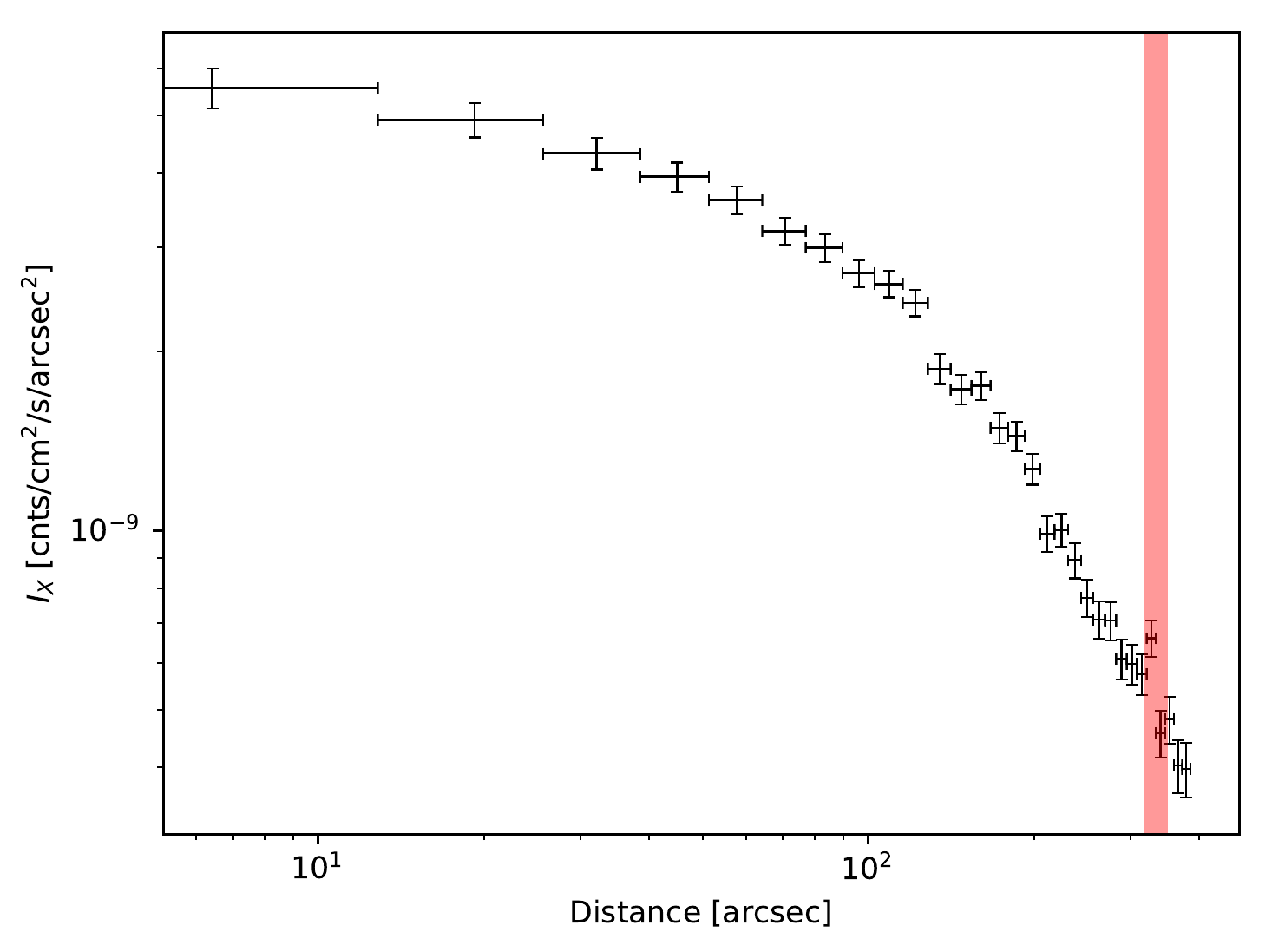}
\includegraphics[width=0.49\textwidth]{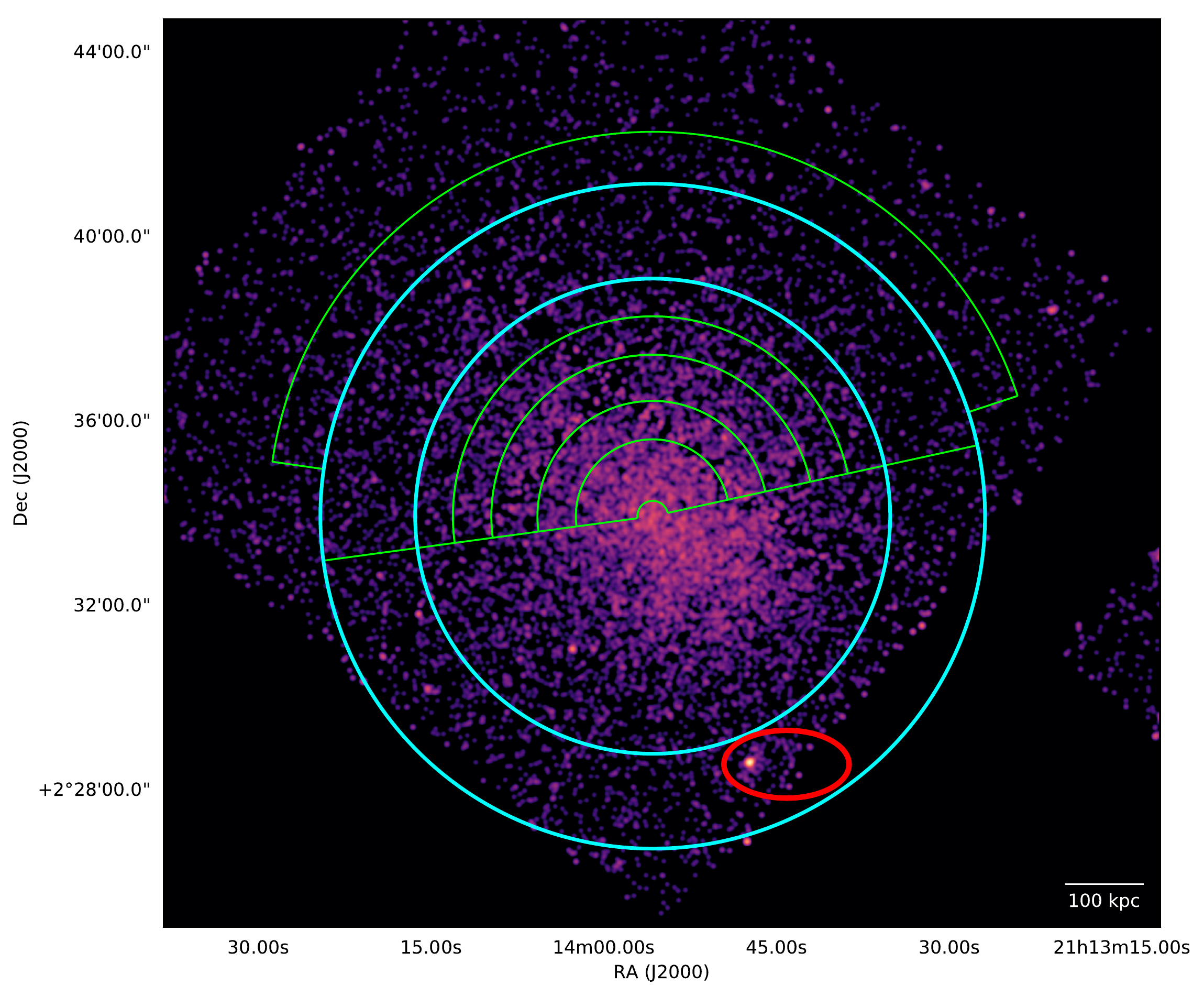}\par
    \caption{\label{fig_X} \textbf{Left}: Background-subtracted, exposure-corrected X-ray surface brightness radial profile in the (0.5--2.0)\,keV energy band, the red band indicates the position of the galaxy \textbf{Right}: Background-subtracted, exposure-corrected image of the cluster in the (0.5--2.0)\,keV band smoothed with a 2.5$^{\prime\prime}$ Gaussian. We show the galaxy region (red), the annulus used for the projected analysis (cyan) and the sectors that we used for the de-projection analysis (green).}
\end{figure*}

In order to estimate the properties of the thermal ICM surrounding JO206, we performed an X-ray analysis of the hosting cluster IIZW108.   
We analysed the archival Chandra observation 10747 (PI: Murray, total exposure time 10.1 ks) of the cluster. The data set was reprocessed with CIAO 4.11 and corrected for known time-dependent gain and charge transfer inefficiency problems following techniques similar to those described in the Chandra analysis threads\footnote{{\ttfamily cxc.harvard.edu/ciao/threads/index.html}}. To filter out strong background flares, we also applied screening of the event files. The remaining, clear exposure time is 10.0\,ks. We used CALDB 4.8.2 blank-sky background files normalised to the count rate of the source image in the (9--12)\,keV band to produce the appropriate background for the observation. We then produced a background-subtracted, exposure-corrected image in the (0.5--2.0)\,keV (Fig. \ref{fig_X}), which shows the cluster emission and a point source located at the galaxy position.\\

\begin{figure}
    \centering
    \includegraphics[width=.5\textwidth]{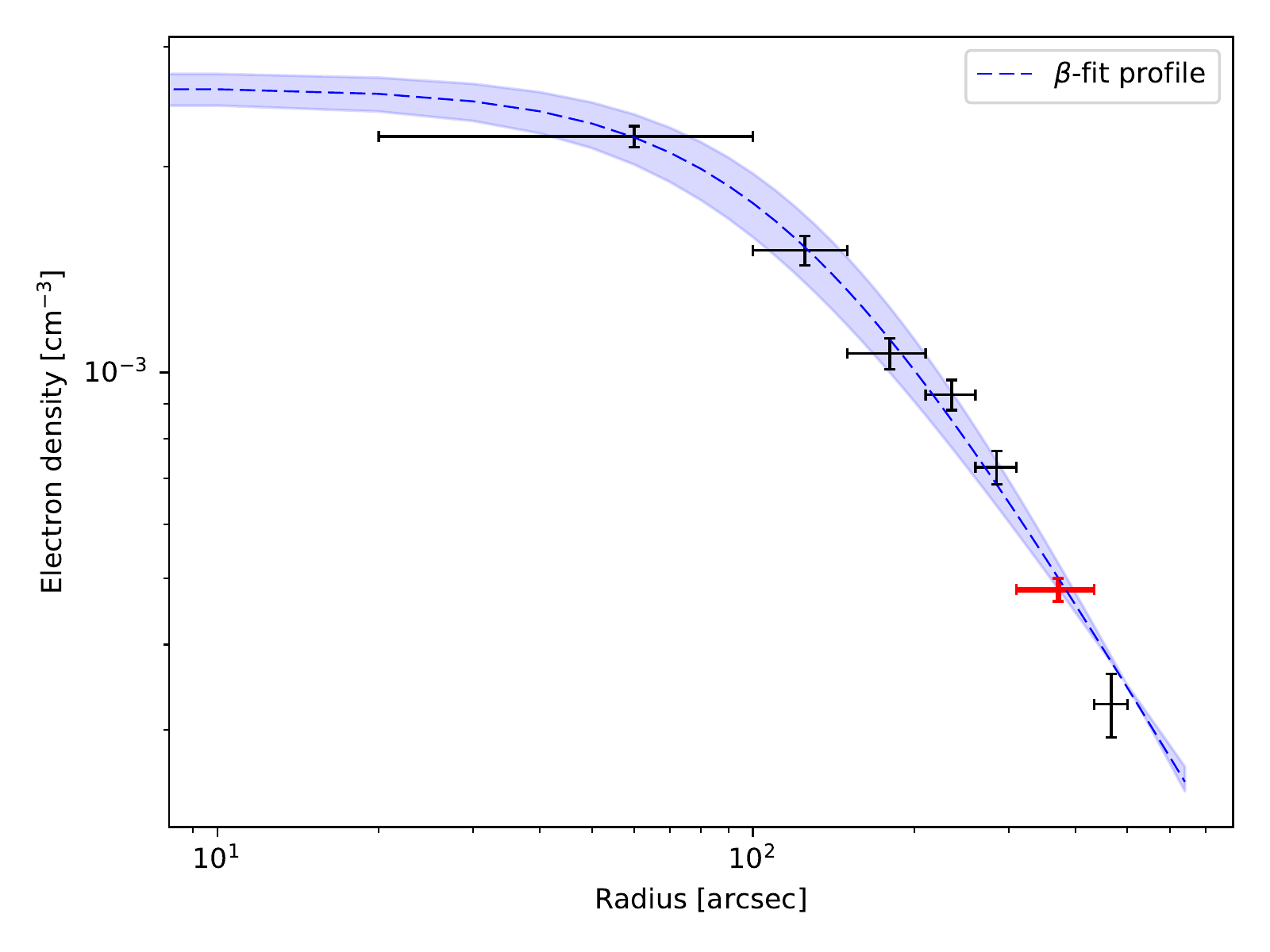}
    \caption{De-projected electron density profile. The red point indicates the sector at the same cluster-centric distance of JO206. In blue we report the best-fit $\beta$-model profile with the 1$\sigma$ confidence interval.}
    \label{dens_x}
\end{figure}

We extracted an azimuthally averaged surface brightness radial profile of the cluster from 30 circular annuli centred on the cluster centre (21:13:55.7738, +2:33:55.400, J2000 reference system) that cover the distance from 15$^{\prime\prime}$ to 400$^{\prime\prime}$. The galaxy is located at $\sim$ 300$^{\prime\prime}$ from the cluster centre and it can be recognised in the profile (Fig. \ref{fig_X}, left). \\
Then we performed a spectral analysis to measure the local properties of the ICM around JO206. We used the cosmological reference of the cluster ($z=0.0482$, angular distance $D_\textrm{A}=201$\,Mpc and 1$^{\prime\prime}=0.976$\,kpc) and we report the error at 1\,$\sigma$. In order to obtain a reliable estimate of the ICM temperature and electron density, a de-projection analysis is required. Unfortunately, the position of the galaxy at the border of the ACIS-I CCD does not allow to perform a de-projection analysis in the direction of the galaxy. However, under the hypothesis of spherical symmetry, we can assume that the thermal plasma has the same radial properties in every direction and, thus, we can estimate the local ICM properties around JO206 by studying the X-ray emission at the same cluster-central distance of the galaxy. Therefore, we extracted the spectra in the 7 semi-circular regions reported in green in Figure~\ref{fig_X}, right panel. As a sanity check, we extracted also a spectrum in the cyan annulus by excluding the galaxy (red ellipse) and we compared its projected temperature with the value estimated in the sixth sector, which is at the same cluster-central distance. We fitted an absorbed thermal emission model ({\ttfamily phabs*apec}) in the (0.5--7.0)\,keV band with XSPEC 12.10.0 \cite{ArnaudSI}. We fixed the total galactic HI column density\footnote{{\ttfamily heasarc.gsfc.nasa.gov/cgi-bin/Tools/w3nh/w3nh.pl}} to $5.32\times 10^{-20}$\,cm$^{-2}$ and the local metallicity to 0.4. We measured consistent temperature values of $kT=3.9_{-0.4}^{+0.4}$\,keV and $kT=4.0_{-0.5}^{+0.6}$\,keV in the cyan and green regions, respectively, confirming our hypothesis of spherical symmetry of the cluster. Then, we fitted the spectra extracted in the green sectors with a projected, absorbed thermal emission model ({\ttfamily projct*phabs*apec}) in the same energy band assuming the same HI column density.  The final $\chi$ statistic is 327.06 with 387 degrees of freedom. The normalisation of the thermal emission model is defined by
\begin{equation}
 N=\frac{10^{-14}}{4\pi\left[D_\textrm{A}(1+z)\right]^2}\int n_\textrm{e} n_\textrm{H}\, \textrm{d}V,
 \label{norm}
\end{equation}
where $n_\textrm{e}$ and $n_\textrm{H}$ are the electron and proton density and $V$ is the emitting volume. Under the hypothesis of spherical symmetry, i.e. by assuming that the emission comes from a spherical shell with the same width of the semi-circular sector, Equation \ref{norm} can be inverted to estimate the electron density in each bin and, thus, we could estimate the electron density radial profile for the whole cluster up to 585\,kpc (Fig. \ref{dens_x}). 
 At the galaxy distance, we measured a de-projected temperature $kT=3.2^{+0.6}_{-0.5}$\,keV and a normalisation $N=(0.0032\pm0.0002)$\,cm$^{-5}$., that corresponds to an ICM electron density of $n_\textrm{e}=(4.9\pm0.2)\times10^{-4}$\,cm$^{-3}$.  This entails a local ICM thermal pressure $P=1.83\, n_\textrm{e}kT=4.6\times10^{-12}$\,erg\,cm$^{-3}$ and a gas density $\rho=1.9\, \mu n_\textrm{e} m_\textrm{H}=9.5\times10^{-28}$\,g\,cm$^{-3}$, where $\mu\simeq0.61$ and $m_\textrm{H}$ are the mean molecular weight and the proton rest mass, respectively.\\
From the spectral analysis we can constrain the galactic dynamic in the cluster. The ICM temperature results in a local speed of sound $c_\textrm{s}=\sqrt{\gamma P/\rho}\simeq$ 910\,km\,s$^{-1}$, where $\gamma=5/3$ is the adiabatic index, $P$ the thermal pressure and $\rho$ the gas density. From MUSE analysis, a line-of-sight velocity of $v_\parallel \sim850$\,km\,s$^{-1}$ is measured \cite{Poggianti17nSI}, but the extended tail on the plane of the sky suggests that the galaxy motion has a significant velocity component perpendicular to the line-of-sight. Thus, by assuming that the transversal velocity is, at least, equal to the line-of-sight velocity, we can derive a putative lower limit of the total galactic velocity of $v_t\geq\sqrt{2}\cdot v_\parallel=1200$\,km\,s$^{-1}$. This implies that the galaxy would be moving supersonically in the ICM with a Mach number M $\geq$1.3. 

Finally, we derived a mathematical function, $n_\textrm{e}(r)$, to estimate the electron density in the cluster. We fitted the observed density profile with a $\beta$-model profile (Cavaliere-Fusco Femiano 1976):
\begin{equation}
    n_\textrm{e}(r)=n_0\left[1+\left(\frac{r}{r_\textrm{c}} \right)^2 \right]^{-3\beta/2},
\end{equation}
where $n_0$ is the central density, $r_\textrm{c}$ is the core radius and $\beta$ describes the ratio between thermal and gravitational energy of the plasma. The best-fit parameters are $n_0=(2.61\pm0.14)\times 10^{-3}$ cm$^{-3}$, $r_\textrm{c}=(112.14\pm21.03)^{\prime\prime}$ and $\beta=0.44\pm0.06$, where the errors are reported with 1\,$\sigma$ confidence and the final coefficient of determination of the fit is $R^2=0.995$.

\end{document}